\documentclass[english,T1]{tlp}
\usepackage{amsmath}
\usepackage{pst-all}
\usepackage{stmaryrd}
\usepackage{graphicx}
\usepackage{amssymb}
\usepackage{fancyvrb}
\usepackage{color}
\usepackage{babel}
\newcommand{\plusplus}{\mathrel+\joinrel\mathrel+}
\newcommand{\avar}{\ensuremath{\mathrel\_\joinrel\mathrel\_}}
\newcommand{\omegap}{\ensuremath{\omega_p}}
\newcommand{\omegat}{\ensuremath{\omega_t}}

\newcommand{\chrrp}{CHR$^{\textrm{rp}}$}
\newcommand{\comment}[1]{}

\newcommand\vrbprime{\ensuremath{^\prime}}

\newcommand\ith[1]{\ensuremath{{#1}^\mathrm{th}}}
\newcommand\vrbsubt[1]{\ensuremath{_{\texttt{#1}}}}
\newcommand\trans[2]{\stackrel{#1}{\rightarrowtail}_{#2}}
\newcommand\ltstrans[1]{\stackrel{#1}{\rightarrow}}
\newcommand\deriv[2]{\ \protect{\stackrel{#1}{\rightarrowtail}{\!\!}_{#2}^*}\ }

\newcommand\norm[1]{\left\Vert{#1}\right\Vert}
\newcommand\partmgu{\protect{\textsf{partition\_to\_mgu}}}
\newcommand\notrans[2]{\stackrel{#1}{\,\not\!\rightarrowtail}_{#2}}
\newcommand\qed{}
\newcommand\fchr{\ensuremath{\textsf{chr}}}
\newcommand\fid{\ensuremath{\mathsf{id}}}
\newcommand\fsplit{\ensuremath{\textsf{split}}}
\newcommand\ffilter{\ensuremath{\textsf{filter}}}
\newcommand\fmodes{\ensuremath{\textsf{modes}}}
\newcommand\fsubtopar{\ensuremath{\textsf{substitution\_to\_partition}}}
\newcommand\chrtola{\ensuremath{\textsf{chr\_to\_la}}}
\newcommand\latochr{\ensuremath{\textsf{la\_to\_chr}}}

\newcommand\true{\ensuremath{\textit{true}}}
\newcommand\token{\ensuremath{\textit{token}}}
\newcommand\del{\ensuremath{\textit{del}}}
\newcommand\varid{\ensuremath{\textit{Id}}}
\newcommand\varids{\ensuremath{\textit{Ids}}}
\newtheorem{definition}{Definition}
\newtheorem{example}{Example}
\newtheorem{theorem}{Theorem}
\newtheorem{lemma}{Lemma}
\newtheorem{corollary}{Corollary}

\newcommand\headmatch{\ensuremath{\mathsf{head\_match}}}

\begin{document}
\VerbatimFootnotes
\title[Logical Algorithms meets Constraint Handling Rules]
{Logical Algorithms meets CHR\\
{\normalsize A Meta-Complexity Result for Constraint Handling Rules with Rule Priorities}}

\author[Leslie De Koninck]
{Leslie De Koninck\\
Department of Computer Science, K.U.Leuven, Belgium\\ 
\email{\textit{FirstName.LastName}@cs.kuleuven.be}}
\submitted{20 December 2007}
\revised{20 December 2008}
\accepted{9 January 2009}
\maketitle
\begin{abstract}
This paper investigates the relationship between the Logical Algorithms
language (LA) of Ganzinger and McAllester and Constraint Handling Rules (CHR).
We present a translation schema from LA to \chrrp{}: CHR with rule priorities,
and show that the meta-complexity theorem for LA can be applied to a subset of
\chrrp{} via inverse translation. Inspired by the high-level implementation
proposal for Logical Algorithm by Ganzinger and McAllester and based on a new
scheduling algorithm, we propose an alternative implementation for \chrrp{} 
that gives strong complexity guarantees and results in a new and accurate 
meta-complexity theorem for \chrrp{}. It is furthermore shown that the 
translation from Logical Algorithms to \chrrp{} combined with the new \chrrp{}
implementation, satisfies the required complexity for the Logical Algorithms
meta-complexity result to hold.
\end{abstract}
\section{Introduction}
Constraint Handling Rules (CHR) \cite{chr} is a high-level rule based language,
originally designed for the implementation of constraint solvers, but also
increasingly used as a general purpose programming language. Recently, it was
shown that all algorithms can be implemented in CHR while preserving both time
and space complexity \cite{jon:complexity}. We assume some familiarity with CHR
and refer to \cite{chr} for more details.

In ``Logical Algorithms'' (LA) \cite{la} (and based on previous work in 
\cite{la2,la1}), Ganzinger and McAllester present a bottom-up logic programming
language for the purpose of facilitating the derivation of complexity results
of algorithms described by logical inference rules. This problem is far from
trivial because the runtime is not necessarily proportional to the derivation
length (i.e., the number of rule applications), but also includes the cost of
pattern matching for multi-headed rules, as well as the costs related to 
high-level execution control which is specified using rule priorities in
the Logical Algorithms language. The language of Ganzinger and McAllester
resembles CHR 
in many ways and has often been referred to
in the discussion of complexity results of CHR programs
\cite{dcg,atgb2,tom:uf,jon:dijkstra}. In particular, in \cite{dcg}, 
Christiansen uses the meta-complexity theorem that accompanies the Logical 
Algorithms language, and notes that the CHR system used (SICStus CHR by
Holzbaur et al. \cite{sicstus_chr}) does not always exhibit the right 
complexity because previously computed partial rule matches are not stored. 

The aim of this paper is to
investigate the relationship between both languages. More precisely, we 
look at how the meta-complexity theorem for Logical Algorithms can be applied
to (a subset of) CHR, and how CHR can be used to implement Logical Algorithms
with the correct complexity.
First, we present a translation schema from Logical Algorithms to \chrrp{}: CHR
extended with rule priorities \cite{chrrp}. Logical Algorithms derivations of
the original program correspond to \chrrp{} derivations in the translation and
vice versa. We also show how to translate a subclass of \chrrp{} programs into
Logical Algorithms. This allows us to apply the meta-complexity theorem for
Logical Algorithms to these \chrrp{} programs as well. Because the 
Logical Algorithms meta-complexity theorem is based on an optimized 
implementation, it gives more accurate results than the implementation 
independent meta-complexity theorem of \cite{atgb1,atgb2}
while being more general than the ad-hoc complexity derivations in
\cite{tom:uf,jon:dijkstra}.

Our current implementation of \chrrp{} as presented in \cite{compiling_chrrp}
does not guarantee the complexity required for the meta-complexity theorem for
Logical Algorithms to hold via translation to \chrrp{}.
Another issue is that the translation from \chrrp{} to Logical 
Algorithms is restricted to a subset of \chrrp{}. Therefore, we propose a new 
implementation of \chrrp{}, designed such that it supports a new 
meta-complexity theorem for the complete \chrrp{} language, while also ensuring
that Logical Algorithms programs translated into \chrrp{} are executed with the
correct complexity. 

The implementation is
based on the high-level implementation proposal for Logical Algorithms as given
in \cite{la}, and on a new scheduling data structure proposed in 
\cite{mergeable_schedules}. By using a CHR system with advanced indexing 
support, such as the K.U.Leuven CHR system  \cite{kulchr}, our implementation 
achieves the complexity required to enable a new and accurate meta-complexity
result for the whole \chrrp{} language.

\paragraph{Overview}
The rest of this paper is organized as follows. In 
Section~\ref{sec:la_and_chr},  
the syntax and semantics of the Logical Algorithms language and \chrrp{} are
reviewed and the known meta-complexity theorems for both languages are
presented. In Section~\ref{la-to-chr} a translation of LA programs to \chrrp{}
programs is presented and in Section~\ref{chr-to-la}, the opposite is done for
a subset of \chrrp{}. Section~\ref{sec:new_implementation} proposes an 
alternative implementation for \chrrp{} which enables a new meta-complexity
theorem for this language, given in Section~\ref{sec:new_meta}. 
Some concluding remarks are given in 
Section~\ref{sec:conclusions}.
\section{Logical Algorithms and \chrrp}\label{sec:la_and_chr}
In this section, we give an overview of the syntax and semantics of Logical
Algorithms (Section~\ref{sec:la_and_chr:la}) and \chrrp{} 
(Section~\ref{sec:la_and_chr:chrrp}). In Section~\ref{sec:la_and_chr:meta}, we
review the meta-complexity results that are known for both languages.
\subsection{Logical Algorithms}\label{sec:la_and_chr:la}
This subsection gives an overview of the syntax and semantics of the Logical
Algorithms language. 
\subsubsection{Syntax}
A Logical Algorithms program $P=\{r_1,\ldots,r_n\}$ is a set of rules.
In \cite{la}, a graphical notation is used to represent rules. We use a new
textual representation that is closer to the syntax of CHR.
A Logical Algorithms rule is an expression 
\[r\ @\ p: A_1,\ldots, A_n\Rightarrow C\]
where $r$ is the rule \emph{name}, the atoms $A_i$ (for $1\leq i\leq n$) are
the \emph{antecedents} and $C$ is the \emph{conclusion}, which is a conjunction
of atoms whose variables appear in the antecedents. Rule $r$ has
\emph{priority} $p$ where $p$ is an arithmetic expression whose variables (if
any) occur in the first antecedent $A_1$. If $p$ contains variables, then $r$
is called a dynamic priority rule. Otherwise, it is called a static priority
rule. In the graphical notation of \cite{la}, the above rule is represented as
shown below.
\[
\begin{array}{ll}
 & A_1\\
 & \:\:\vdots\\
 & A_n\\
\textrm{(r,p)} & ^{\underline{\quad\:\:}}\\
 & C\end{array}\]
The arguments of an atom are either Herbrand terms or (integer) arithmetic
expressions. There are two types of atoms: comparisons and user-defined atoms.
A comparison has the form $x<y$, $x\leq y$, $x=y$ or $x\neq y$ with $x$ and $y$
arithmetic expressions or, in case of $(=)/2$ and $(\neq)/2$, Herbrand terms. 
Comparisons are only allowed in the antecedents of a rule and all variables in
a comparison must appear in earlier antecedents.
A user-defined atom can be positive or negative. A negative user-defined atom
has the form $del(A)$ where $A$ is a positive user-defined atom. 
A ground user-defined atom is called an assertion.
\begin{example}
An example rule (from Dijkstra's shortest path algorithm as presented in 
\cite{la}) with name \texttt{d2} and priority 1 is
\begin{Verbatim}[commandchars=*\{\},frame=single,fontsize=\small]
d2 @ 1 : dist(V,D*vrbsubt{1}), dist(V,D*vrbsubt{2}), D*vrbsubt{2} < D*vrbsubt{1} => del(dist(V,D*vrbsubt{1})).
\end{Verbatim}
The antecedent \texttt{D\vrbsubt{2} < D\vrbsubt{1}} is a comparison, the atoms
\texttt{dist(V,D\vrbsubt{1})} and \texttt{dist(V,D\vrbsubt{2})} are positive
user-defined antecedents. The negative 
ground atom \texttt{del(dist(a,5))} is an example of a negative assertion.
\end{example}
\subsubsection{Operational Semantics}
A Logical Algorithms state $\sigma$ consists of a set of (positive and 
negative) assertions. A state can simultaneously contain the positive assertion
$A$ and the negative assertion $del(A)$. In the rest of this paper, we
sometimes use the word \emph{database} as a synonym for a LA execution state.
Let $\mathcal{D}$ be the usual interpretation for the comparisons. 
Given a program $P$, the following transition converts
one state into the next:
\begin{center}
\fbox{
\begin{minipage}[c]{0.95\textwidth}
\begin{description}
\item[1. Apply] 
$\sigma\stackrel{LA}{\rightarrowtail}_P\sigma\cup\theta(C)$
if there exists a (renamed apart) rule $r$ in $P$ of priority $p$ of the
form \[r\ @\ p:A_1,\ldots, A_n \Rightarrow C\]
and a ground substitution $\theta$ such that for every antecedent 
$A_i$,
\begin{itemize}
\item $\mathcal{D}\models \theta(A_i)$ if $A_i$ is a comparison
\item $\theta(A_i)\in\sigma$ and $del(\theta(A_i))\notin\sigma$
    if $A_i$ is a positive user-defined atom
\item $\theta(A_i)\in\sigma$ if $A_i$ is a negative user-defined atom 
\end{itemize}
Furthermore, $\theta(C)\nsubseteq\sigma$ and no rule of priority $p'$ and
substitution $\theta'$ exists with $\theta'(p')<\theta(p)$ for which the above
conditions hold.
\end{description}
\end{minipage}
}
\end{center}
A state is called final if no more transitions apply to it. A non-final state
has priority $p$ if the next firing rule instance has priority $p$. The
condition $\theta(C)\nsubseteq\sigma$ ensures that no rule instance fires more
than once and prevents trivial non-termination. This condition, combined with
the fact that each transition only creates new assertions, causes
the consecutive states in a derivation to be monotone increasing.
Although the priorities restrict the possible derivations, the choice of which
rule instance to fire from those with equal priority is non-deterministic.
\subsection{\chrrp: CHR with Rule Priorities}\label{sec:la_and_chr:chrrp}
\chrrp{} is CHR extended with user-definable rule priorities. It is introduced
in \cite{chrrp} as a solution to the lack of high-level execution control in
CHR. In this section, we review the syntax and semantics of \chrrp{}.
\subsubsection{Syntax}\label{sec:la_and_chr:chrrp:syntax}
A constraint $c(t_1,\ldots,t_n)$ is an atom of predicate $c/n$ with $t_i$ a
host language value (e.g., a Herbrand term in Prolog) for $1\leq i\leq n$. There are two types of constraints: 
built-in constraints and CHR constraints (also called user-defined
constraints). The CHR constraints are solved by the CHR program whereas the
built-in constraints are solved by an underlying constraint solver (e.g., the
Prolog unification algorithm).

There are three types of Constraint Handling Rules: 
\emph{simplification rules}, \emph{propagation rules} and 
\emph{simpagation rules}. They have the following form:
\[\begin{array}{ll@{\ }l@{\ }r@{\ }c@{\ }l}
\textbf{Simplification}&p::r\ @&&H^{r}&\iff&g\mid B\\
\textbf{Propagation}&p::r\ @&H^{k}&&\implies&g\mid B\\
\textbf{Simpagation}&p::r\ @&H^{k}&\backslash\ H^{r}&\iff&g\mid B\end{array}\]
where $p$ is the rule priority, $r$ is the rule \emph{name}, $H^k$ and $H^r$ 
are non-empty sequences of CHR constraints and are called the \emph{heads} of
the rule. The rule \emph{guard} $g$ is a sequence of built-in constraints and
the rule \emph{body} $B$ is a sequence of both CHR and built-in constraints.
The rule priority is either a number in which case the rule is called a
\emph{static} priority rule, or an arithmetic expression whose variables appear
in the heads $H^k$ and/or $H^r$ in which case the rule is called a \emph{dynamic}
priority rule. We say that priority $p$ is higher than priority $p'$ if $p<p'$.
For simplicity, we sometimes assume priorities are integers and the highest
priority is 1. Finally, a program $P$ is a set of CHR rules. Apart from the
rule priorities, \chrrp\ is identical to CHR.
\subsubsection{Operational Semantics}\label{sec:la_and_chr:chrrp:semantics}
Operationally, CHR constraints have a multi-set semantics. To distinguish
between different occurrences of syntactically equal constraints, CHR 
constraints are extended with a uni\-que identifier. An identified CHR
constraint
is denoted by $c\#i$ with $c$ a CHR constraint and $i$ the identifier. We write
$\fchr(c\#i)=c$ and $\fid(c\#i)=i$ and pointwise extend these functions to sets
and sequences of constraints.

The operational semantics of \chrrp{}, called the priority semantics and
denoted by \omegap{}, is given in \cite{chrrp} as a state transition system, 
similar to the approach of \cite{duck:refined} for the theoretical and refined
operational semantics of CHR. A CHR execution state $\sigma$ is represented as
a tuple $\langle G,S,B,T\rangle_n$ where $G$ is the goal, a multi-set of 
constraints to be solved; $S$ is the CHR constraint store, a set of
identified CHR constraints; $B$ is the built-in store, a conjunction
of built-in constraints; $T$ is the propagation history, a set of tuples
denoting the rule instances that have already fired; and $n$ is the next free
identifier, used to identify new CHR constraints. The transitions of \omegap{}
are shown in Table \ref{tab:omegap} where $\mathcal{D}$ denotes the built-in
constraint theory and $\bar{\exists}_X Y$ denotes the existential closure of
$Y$ apart from the variables in $X$. The transitions are exhaustively applied
starting from the state $\langle G,\emptyset,true,\emptyset\rangle_1$
with $G$ the initial goal. We have used the simpagation rule form to denote any
type of rule in the \textbf{Apply} transition. For simplification rules, $H_1$
and $H'_1$ are empty, and for propagation rules, $H_2$ and $H'_2$ are empty.
\begin{table}
\fbox{
\begin{minipage}[l]{0.95\linewidth}
\begin{description}
\item[1. Solve] $\langle\{c\}\uplus G,S,B,T\rangle_n\trans{\omegap}{P}
\langle G,S,c\wedge B,T\rangle_n$ where $c$ is a built-in constraint.
\item[2. Introduce] $\langle\{c\}\uplus G,S,B,T\rangle_n\trans{\omegap}{P}
\langle G,\{c\#n\}\cup S,B,T\rangle_{n+1}$ where $c$ is a CHR constraint.
\item[3. Apply] $\langle\emptyset,H_1\cup H_2\cup S,B,T\rangle_n
\trans{\omegap}{P}
\langle \theta(C),H_1\cup S,B,T\cup\{t\}\rangle_n$ where
$P$ contains a rule of priority $p$ of the form 
\[p::r\ @\ H'_1\backslash H'_2 \iff g\mid C\]
and a matching substitution $\theta$ such that $\fchr(H_1)=\theta(H'_1)$,
$\fchr(H_2)=\theta(H'_2)$, $\mathcal{D}\models B\rightarrow\bar{\exists}_B
\theta(g)$; $\theta(p)$ is a ground arithmetic expression and 
$t= \langle r, \fid(H_1)\plusplus \fid(H_2)\rangle\notin 
T$. Furthermore, no rule of priority $p'$ and substitution $\theta'$ exists
with $\theta'(p')<\theta(p)$ for which the above conditions hold. 
\end{description}
\end{minipage}}
\caption{Transitions of \omegap}\label{tab:omegap}
\end{table}

The following theorem on the correspondence between the \omegap\ semantics of
\chrrp\ and the theoretical operational semantics \omegat{} of CHR (see e.g.
\cite{duck:refined}), is proven in \cite{chrrp}.
\begin{theorem}
Every derivation $D$ under \omegap\ is also a derivation under \omegat. If a
state $\sigma$ is a final state under \omegap, then it is also a final state
under \omegat.
\end{theorem}
In the refined operational semantics of CHR \cite{duck:refined}, the textual
order of the program rules determines which rule is tried next for the
current \emph{active} constraint. However, only rule instances in which the
active constraint takes part are considered, and so a higher priority fireable
rule instance in which the active constraint does not participate, will not 
fire. The textual rule order also does not support dynamic rule priorities.
\subsubsection{Differences compared to Logical Algorithms}
\label{sec:la_and_chr:chr:differences}
\chrrp{} differs from Logical Algorithms in the following ways:
\begin{itemize}
\item A Logical Algorithms state is a set of ground assertions, while the CHR
constraint store is a multi-set and may also contain non-ground constraints.
\item In Logical Algorithms, built-in constraints are restricted to ask
constraints and only include comparisons; \chrrp{} 
supports any kind of built-in constraints.
\item A removed CHR constraint may be reasserted and can then participate again
in rule firings whereas a removed LA assertion cannot be asserted again.
\item A Logical Algorithms rule may contain negated heads. In contrast, 
\chrrp{} requires all heads to be positive.%
\footnote{See \cite{peter:negation} for an extension of CHR with negation as
absence. However, the semantics of that form of negation is different from the
one in Logical Algorithms.}
\item In the Logical Algorithms language, the priority of a dynamic priority
rule is determined by the variables in the left-most head, whereas in \chrrp{}
it may depend on multiple heads.
\end{itemize}
We note that rules
for which the priority depends on more than one head, can easily be transformed
into the correct form as follows. Given a Logical Algorithms rule of the form
\[r\ @\ p:A_1,\ldots,A_m,A_{m+1},\ldots,A_n\Rightarrow C\]
where the priority expression $p$ is fully determined by the variables from the
antecedents $A_1,\ldots,A_m$. This rule can be transformed into the equivalent
rules
\begin{align*}
r_1\ @\ 1 &: A_1,\ldots,A_m\Rightarrow \textsf{priority}_r(p)\\
r_2\ @\ p &: \textsf{priority}_r(p), A_1,\ldots,A_m,A_{m+1},\ldots, A_n\Rightarrow C 
\end{align*}
where $\textsf{priority}_r$/1 is a new user-defined predicate. Now the first
head of the dynamic priority rule determines the rule priority. The above 
transformation causes the creation of $\textsf{priority}_r$/1 assertions.
We have that every execution state of the transformed program can be mapped on
a corresponding execution state of the original program (assuming rule
priorities are allowed to depend on multiple heads) by removing these 
$\textsf{priority}_r$/1 assertions.
\subsection{Meta-Complexity Results}\label{sec:la_and_chr:meta}
The Logical Algorithms language was designed with a meta-complexity result in
mind. Such a result has also been formulated for CHR. In this subsection, we
review both results and give a first intuition on how they relate to each 
other.
\subsubsection{The Logical Algorithms Meta-Complexity Result}
\label{sec:la_and_chr:meta:la}
A \emph{prefix instance} of a Logical Algorithms rule 
$r\ @\ p:A_1,\ldots, A_n \Rightarrow C$ is a tuple
$\langle r,i,\theta\rangle$ with $\theta$ a ground substitution defined on the
variables occuring in $A_1,\ldots,A_i$ and $1\leq i\leq n$. Its antecedents are
$\theta(A_1),\ldots,\theta(A_i)$. A \emph{strong prefix firing} is a prefix 
instance whose antecedents hold in a state with priority lower or equal to the
prefix' rule priority.
In \cite{la}, also the concept of a \emph{weak} prefix firing is
defined, but it is of no importance for our purposes.
The time complexity for running Logical Algorithms programs is given in
\cite{la} as $\mathcal{O}(|\sigma_0|+P_s+(P_d+A_d)\cdot\log N)$ where 
$\sigma_0$ is the initial state and $|\sigma_0|$ is its size. $P_s$ is the
number of strong prefix firings of static priority rules and $P_d$ is
the number of strong prefix firings of dynamic priority rules;
$A_d$ is the
number of assertions that may participate in a dynamic priority rule instance;
and $N$ is the number of distinct priorities. The following example is
adapted from \cite{la}.
\begin{example}[Dijkstra's Shortest Path]\label{ex:dijkstra}
The rules below implement Dijkstra's single source shortest path algorithm. 
\begin{Verbatim}[commandchars=*\{\},frame=single,fontsize=\small]
d1 @ 1   : source(V) => dist(V,0).
d2 @ 1   : dist(V,D*vrbsubt{1}), dist(V,D*vrbsubt{2}), D*vrbsubt{2} < D*vrbsubt{1} => del(dist(V,D*vrbsubt{1})).
d3 @ D+2 : dist(V,D), e(V,C,U) => dist(U,D+C).
\end{Verbatim}
A \texttt{source($V$)} assertion means that $V$ is the (unique) source node for
the algorithm. A \texttt{dist($V$,$D$)} assertion means that the shortest path
distance from the source node to node $V$ does not exceed $D$. Finally, an 
\texttt{e($V$,$C$,$U$)} assertion means that there is an edge from node $V$ to
node $U$ with cost (weight) $C$. Given an initial state consisting of one
\texttt{source}/1 assertion and $e$ \texttt{e}/3 assertions, we can derive that
the 
number of strong prefix firings is $\mathcal{O}(1)$ for rule \texttt{d1}, and 
$\mathcal{O}(e)$ for both rules \texttt{d2} and \texttt{d3}, and so both $P_s$
and $P_d$ are $\mathcal{O}(e)$. This result
is based on the fact that at priority 2 and lower (numerically larger), there
is at most one
(positive) \texttt{dist}/2 assertion for each node, and each of these
assertions 
represent the shortest path distance from the source node to this node. This
means that at most $e$ \texttt{dist}/2 assertions are ever created, and 
$A_d=\mathcal{O}(e)$. Finally, the number of distinct priorities is bounded by
the number of \texttt{dist}/2 assertions, i.e., $N=\mathcal{O}(e)$.
Using the meta-complexity theorem, we find that the total complexity is
$\mathcal{O}(e+e+(e+e)\cdot\log e)=\mathcal{O}(e\log e)$.
\end{example}
\subsubsection{The ``As Time Goes By'' Approach}
\label{sec:la_and_chr:meta:atgb}
In \cite{atgb1,atgb2}, an upper bound on the worst case time complexity of a
CHR program $P$ is given as
\begin{equation}\label{eq:atgb}
\mathcal{O}\left(D\sum_{r\in P}\left(c^{n_r}_{max}\left(O_{H_r}+O_{G_r}
    \right)+\left(O_{C_r}+O_{B_r}\right)\right)\right)
\end{equation}
where $D$ is the maximal derivation length (i.e., the maximal number of rule
firings), $c_{max}$ is the maximal number of CHR constraints in the store,
and for each rule $r\in P$:
\begin{itemize}
\item $n_r$ is the number of heads in $r$
\item $O_{H_r}$ is the cost of head matching, i.e. checking that a given 
sequence of $n_r$ constraints match with the $n_r$ heads of rule $r$
\item $O_{G_r}$ is the cost of checking the guard
\item $O_{C_r}$ is the cost of adding built-in constraints after firing
\item $O_{B_r}$ is the cost of adding and removing CHR constraints after firing
\end{itemize}
For programs with simplification and simpagation rules only, the maximal 
derivation length can be derived using an appropriate ranking on constraints 
that decreases after each rule firing \cite{termination}. We note that finding
such a ranking is not trivial. The meta-complexity result is based on a 
naive CHR implementation, and therefore on the one hand gives an upper bound on
the time complexity for any reasonable implementation of CHR, but on the other
hand often largely overestimates the worst case time complexity on optimized 
implementations.\footnote{Built-in constraints may lead to a worse complexity
in practical optimized implementations if many constraints are repeatedly
reactivated without this resulting in new rule firings. We return to this issue
in Section \ref{sec:new_meta:comparison_atgb}.} 
The following example is adapted from \cite{atgb1}.
\begin{example}[Boolean]
The rules below implement the boolean $and(X,Y,X\wedge Y)$ constraint given 
that 1 represents \emph{true} and 0 represents \emph{false}.
\begin{Verbatim}[frame=single,fontsize=\small]
and(0,Y,Z) <=> Z = 0.                 and(X,0,Z) <=> Z = 0.
and(X,1,Z) <=> X = Z.                 and(1,Y,Z) <=> Y = Z.
and(X,X,Z) <=> X = Z.                 and(X,Y,1) <=> X = 1, Y = 1. 
\end{Verbatim}
Let the rank of an \verb!and!/3 constraint be one, then the rank of the head
of each rule equals one, and the rank of the body equals zero because by
definition, all built-in constraints have a rank of zero.
For a goal consisting of $n$ \verb!and!/3 constraints, the derivation length
is $n$, which is also the maximal number of CHR constraints in the store. The
cost of head matching, (implicit) guard checking, removing CHR constraints and
asserting built-in constraints can all be considered constant. Then using 
\eqref{eq:atgb}, we derive that the total runtime complexity is 
$\mathcal{O}(n^2)$.
\end{example}
\subsubsection{A First Comparison}\label{sec:la_and_chr:meta:comparison}
Although at this point we do not intend to make a complete comparison between
both results, we can already show that the Logical Algorithms result in a sense
is at least as accurate as Fr\"uhwirth's approach, at least as far as programs
without built-in tell constraints are concerned. The reasoning is as follows.
In each derivation step, a constant number of atoms (constraints) are asserted.
Let $c_{max}$ be the maximal number of (strictly) positive assertions in any
given state. Furthermore assume rules have positive heads only,
then each of the asserted atoms can participate in at most 
$\sum_{r\in P}\left(n_r\cdot c^{n_r-1}_{max}\right)$ strong prefix firings. 
Because only $\mathcal{O}(c+D)$ constraints are ever asserted where
$c$ is the number of CHR constraints in the initial goal and $D$ is the 
derivation length, the total number of strong prefix firings $P_s+P_d$ is
\[
\mathcal{O}\left((c+D)\cdot \sum_{r\in P}c^{n_r-1}_{max}\right)
\]
and because $c=\mathcal{O}(c_{max})$ we also have the following bound
\begin{equation}\label{eq:la_and_chr:strong_prefix_firings}
\mathcal{O}\left(D\cdot \sum_{r\in P}c^{n_r}_{max}\right)
\end{equation}
In absence of (dynamic) priorities, the total runtime complexity according to
the Logical Algorithms meta-complexity result is bounded by the same formula 
\eqref{eq:la_and_chr:strong_prefix_firings} and hence is at least as accurate
as the result of \cite{atgb2} given that the cost of both head matching 
($O_{H_r}$) and adding and removing CHR constraints ($O_{B_r}$) is constant for
each rule $r$.
\section{Translating Logical Algorithms into \chrrp}\label{la-to-chr}
In this section, we show how Logical Algorithms programs can be 
translated into \chrrp{} programs. CHR states of the translated program can be
mapped onto LA states of the original. With respect to this mapping, both 
programs have the same derivations. 
\subsection{The Translation Schema\label{rule-trans}}
The translation of a LA program $P$ is denoted by 
$T(P)=T_\textit{S/D}(P)\cup T_\textit{R}(P)$. The definitions of 
$T_{S/D}(P)$ and $T_{R}(P)$ are given below.
\subsubsection{Set and Deletion Semantics}
\label{sec:la_to_chr:schema:set_and_deletion}
We represent Logical Algorithms assertions as CHR constraints consisting
of the assertion itself and an extra argument, called the \emph{mode
indicator}, denoting whether it is positively asserted (``\texttt{p}''),
negatively asserted (``\texttt{n}'') or both (``\texttt{b}''). For every
user-defined predicate $a/n$ occurring in $P$, $T_{S/D}(P)$ contains the
following rules to deal with a new positive or negative assertion:
\begin{align*}
1 :: a_r(\bar{X},M)\ \backslash\ a(\bar{X}) &\iff M \neq \mathtt{n}\mid true\\
1 :: a_r(\bar{X},\mathtt{n}), a(\bar{X}) &\iff a_r(\bar{X},\mathtt{b})\\
2 :: a(\bar{X})&\iff a_r(\bar{X},\mathtt{p})\\
\vphantom{\stackrel{\displaystyle{m}}{T}}%
1 :: a_r(\bar{X},M)\ \backslash\ del(a(\bar{X}))&\iff M\neq\mathtt{p}\mid true\\
1 :: a_r(\bar{X},\mathtt{p}), del(a(\bar{X}))&\iff a_r(\bar{X},\mathtt{b})\\
2 :: del(a(\bar{X}))&\iff a_r(\bar{X},\mathtt{n})
\end{align*}
If a representation already exists, one of the priority 1 rules updates this
representation. Otherwise, one of the priority 2 rules generates a new
representation. At lower (numerically larger) priorities, it is guaranteed that
every assertion,
whether asserted positively, negatively or both, is represented by 
exactly one constraint in the store.
\subsubsection{Rules}
\label{sec:la_and_chr:schema:rules}
Given a LA rule $r\in P$ of the form
\[r\ @\ p:A_1,\ldots,A_n\Rightarrow C\] 
We first split up the antecedents into user-defined antecedents and comparison
antecedents by using the \fsplit{} function defined below.
\begin{align*}
\fsplit([A|T]) = & 
\begin{cases}
\langle [A|A^u],A^c\rangle&
\textrm{if }A\textrm{ is a user-defined atom}\\
\langle A^u,[A|A^c]\rangle&
\textrm{if }A\textrm{ is a comparison}
\end{cases}\\
&\textrm{where }\fsplit(T)=\langle A^u,A^c\rangle\\
\fsplit([]) = &\langle [],[]\rangle 
\end{align*} 
In the Logical Algorithms language, a given assertion may participate multiple
times in the same rule instance, whereas in CHR all constraints in a single
rule instance must be different. To overcome this semantic difference, a single
LA rule is translated as a set of CHR rules such that every
CHR rule covers a case of syntactically equal head constraints.
Let $\langle A^u,A^c\rangle=\fsplit([A_1,\ldots,A_n])$ with 
$A^u=[A^u_1,\ldots,A^u_m]$ and $A^c=[A^c_1,\ldots,A^c_l]$. Let 
$\mathcal{P}$ be the set of all partitions of $\{1,\ldots,m\}$.
For a given partition $\rho\in\mathcal{P}$, the following function returns
the most general unifier that unifies all antecedents $\{A_i\mid i\in S\}$ for
every $S\in \rho$ where $\mathsf{mgu}(S)$ is the most general unifier of all
elements in $S$.
\[\partmgu(\rho,[A^u_1,\ldots,A^u_m])=
\underset{S\in\rho}{\circ}\mathsf{mgu}(\{A^u_i\mid i\in S\})\]
Let $\mathcal{PU}=\{\langle \rho,\theta\rangle\mid \rho\in\mathcal{P}\wedge
\theta=\partmgu(\rho,A^u)\wedge 
\mathcal{D}\models\bar{\exists}_\emptyset \theta(A^c)\}$. $\mathcal{PU}$ 
contains all partitions for which $\partmgu$ is defined and
for which the comparison antecedents $A^c$ are still satisfiable
after applying the unifier. The next step is to filter out antecedents so that
every set in the partition has only one representative.  
This is done by computing $\ffilter(\theta(A^u),\rho)$
for each $\langle\rho,\theta\rangle\in\mathcal{PU}$
where the \ffilter{} function is as follows:
\begin{align*}
\ffilter([\theta(A^u_i)|T],\rho) = & 
\begin{cases}
[\theta(A^u_i)|\ffilter(T,\rho)]&
\textrm{if }\exists S\in\rho: i=\min(S)\\
\ffilter(T,\rho)&
\textrm{otherwise}
\end{cases}\\
filter([],\avar) = &[]
\end{align*}
Finally, we add mode indicators to all remaining user-defined antecedents:
\begin{align*}
\fmodes([A^{u'}|T]) = &
\begin{cases}
\langle [a_r(\bar{X},\mathtt{p})|A^m],N\rangle&
\textrm{if }A^{u'}=a(\bar{X})\\
\langle [a_r(\bar{X},N')|A^m],[N'\neq\mathtt{p}|N]\rangle&
\textrm{if }A^{u'}=del(a(\bar{X}))
\end{cases}\\
&\textrm{where }\langle A^m,N\rangle=\fmodes(T)\\
\fmodes([]) = &\langle [],[]\rangle
\end{align*}
The \fmodes{} function returns both the resulting antecedents and the
necessary conditions on the mode indicators of these antecedents.
For every $\langle \rho,\theta\rangle\in\mathcal{PU}$, the CHR translation
$T_R(P)$ contains a rule
\[p+2 :: r_{\rho}\ @\ H \implies g_1,g_2\mid C'\]
where $\langle H,g_1\rangle=
    \fmodes(\ffilter(\theta(A^u),\rho))$,
    $g_2=\theta(A^c)$ and $C'=\theta(C)$.
\subsubsection{Examples}
We illustrate the translation schema on some examples.
\begin{example}\label{dijkstra}
A LA implementation of Dijkstra's shortest path algorithm is
\begin{Verbatim}[commandchars=*\{\},frame=single,fontsize=\small]
d1 @ 1   : source(V) => dist(V,0).
d2 @ 1   : dist(V,D*vrbsubt{1}), dist(V,D*vrbsubt{2}), D*vrbsubt{2} < D*vrbsubt{1} => del(dist(V,D*vrbsubt{1})).
d3 @ D+2 : dist(V,D), e(V,C,U) => dist(U,D+C).
\end{Verbatim}
Its translation is
\begin{Verbatim}[commandchars=*\{\},frame=single,fontsize=\small]
1 :: e*vrbsubt{r}(V,C,U,M) \ e(V,C,U) <=> M \= n | true.
1 :: e*vrbsubt{r}(V,C,U,n) , e(V,C,U) <=> e*vrbsubt{r}(V,C,U,b).
2 :: *vrbsubt{* }             e(V,C,U) <=> e*vrbsubt{r}(V,C,U,p).

1 :: e*vrbsubt{r}(V,C,U,M) \ del(e(V,C,U)) <=> M \= p | true.
1 :: e*vrbsubt{r}(V,C,U,p) , del(e(V,C,U)) <=> e*vrbsubt{r}(V,C,U,b).
2 :: *vrbsubt{* }             del(e(V,C,U)) <=> e*vrbsubt{r}(V,C,U,n).

*ldots % *vrb{(similar rules for }source*vrb{/1 and }dist*vrb{/2)}

  3 :: d1*vrbsubt{1* * } @ source*vrbsubt{r}(V,p) ==> dist(V,0).
  3 :: d2*vrbsubt{1/2} @ dist*vrbsubt{r}(V,D*vrbsubt{1},p), dist*vrbsubt{r}(V,D*vrbsubt{2},p) ==> D*vrbsubt{2} < D*vrbsubt{1} | del(dist(V,D*vrbsubt{1})).
D+4 :: d3*vrbsubt{1/2} @ dist*vrbsubt{r}(V,D,p), e*vrbsubt{r}(V,C,U,p) ==> dist(U,D+C).
\end{Verbatim}
\end{example}
\begin{example}
A rule from the union-find implementation of \cite{la} is the following:
\begin{Verbatim}[frame=single,fontsize=\small]
uf4 @ 1 : union(X,Y), find(X,Z), find(Y,Z) => del(union(X,Y)).
\end{Verbatim}
Because antecedents \texttt{find(X,Z)} and \texttt{find(Y,Z)} are unifiable,
this leads to the following two CHR rules:
\begin{Verbatim}[commandchars=*\{\},frame=single,fontsize=\small]
3 :: uf4*vrbsubt{1/2/3} @ union*vrbsubt{r}(X,Y,p),*:find*vrbsubt{r}(X,Z,p),*:find*vrbsubt{r}(Y,Z,p) ==> del(union(X,Y)).
3 :: uf4*vrbsubt{1/23* } @ union*vrbsubt{r}(X,X,p),*:find*vrbsubt{r}(X,Z,p) ==> del(union(X,X)).
\end{Verbatim}
\end{example}

\subsection{The Correspondence between LA and \chrrp\ Derivations\label{corresp}}
In this subsection, we show that every derivation of the original program under
the Logical Algorithms semantics, corresponds to a derivation of the 
translation under the \omegap{} semantics of \chrrp{}. In order to do so, we
introduce a mapping function \chrtola{} between \emph{reachable} CHR execution
states and Logical Algorithms states; see \cite{observable} for a
formal definition of reachability.
Reachability is considered with respect to initial states of the form 
$\langle G,\emptyset,true,\emptyset\rangle_n$ where the user-defined 
constraints in $G$ are of the form $a(\bar{X})$ and $del(a(\bar{X}))$ and do 
not include constraints of the form $a_r(\bar{X},M)$.
\begin{align*}
\chrtola(\sigma)=&\ \{a(\bar{X})\mid a(\bar{X})\in A\vee
    (a_r(\bar{X},M)\in A\wedge M\neq\mathtt{n})\}\\ \cup&
\ \{del(a(\bar{X}))\mid del(a(\bar{X}))\in A\vee
    (a_r(\bar{X},M)\in A\wedge M\neq\mathtt{p})\}
\end{align*}
where $\sigma=\langle G,S,B,T\rangle_n$ and $A=G\cup chr(S)$. The mapping 
function also takes into account the constraints that are still in the goal and
those for which the set and deletion semantics rules have not yet fired.
In the rest of this section, we first show how CHR execution states are 
normalized and then show that in a Logical Algorithms state and its 
corresponding normalized CHR execution state, corresponding rule instances can
fire. We start by defining a \emph{pre-normal form}.
\begin{definition}[Pre-normal Form]
A (reachable) state $\sigma$ is in pre-normal form if and only if
$\sigma=\langle \emptyset,S,\true,T\rangle_n$, all constraints in $S$ are of 
the form $a_r(\bar{X},M)\#i$, and if $a_r(\bar{X},M_1)\#i_1\in S$ and 
$a_r(\bar{X},M_2)\#i_2\in S$ then $i_1=i_2$ (and consequently $M_1=M_2$).
\end{definition}
The following lemma shows that every reachable state is \emph{pre-normalized}
before rules are tried with priority $>2$.
\begin{lemma}[Pre-normalization]
For every reachable state $\sigma$, there exists a finite derivation 
$D=\sigma\deriv{\omegap}{T(P)}\sigma^*$ such that
$\sigma^*$ is in pre-normal form, $\chrtola(\sigma)=\chrtola(\sigma^*)$,
and all rules fired in $D$ have priority 1 or 2. Every state has a unique 
pre-normal form with respect to the \chrtola{} mapping function.
\label{lem:prenorm}
\end{lemma}
\begin{proof}
We introduce the following ranking function on CHR states:
\[\norm{\sigma}=2\cdot \left|\{a(\bar{X})\mid a(\bar{X})\in A\}\uplus
\{del(a(\bar{X}))\mid del(a(\bar{X}))\in A\}\right|\\+\left|G\right|\]
where $\sigma=\langle G,S,true,T\rangle_n$, $A=G\uplus\fchr(S)$ and if
$X$ is a (multi-)set, $\left|X\right|$ is its cardinality. 
Clearly, the rank of any state is positive, and if $\norm{\sigma}=0$, state
$\sigma$ is in pre-normal form. If $\sigma$ is not in pre-normal form, then
there exists at least one transition $\sigma\trans{\omegap}{T(P)}\sigma'$.
We show that for all such transitions $\chrtola(\sigma)=\chrtola(\sigma')$
and $\norm{\sigma'}<\norm{\sigma}$, which ensures termination. 

If the goal $G$ is not empty, then only the \textbf{Introduce} transition is
applicable. Every application of this transition moves a CHR constraint from
the goal to the CHR constraint store, so $\norm{\sigma'}=\norm{\sigma}-1$.
By definition, $\chrtola(\sigma')=\chrtola(\sigma)$ (because the $\chrtola$
function does not 
distinguish between the goal and the CHR constraint store). 

If the goal $G$ is empty then given that $\sigma$ is not in pre-normal form,
$\fchr(S)$ contains a constraint of the form $a(\bar{X})$ or $del(a(\bar{X}))$.
We look into detail to the case of $a(\bar{X})\in\fchr(S)$; the case of
$del(a(\bar{X}))\in\fchr(S)$ is similar. We start by showing that at least one
rule of priority 1 or 2 is applicable. Next, we show that each rule application
decreases the norm and maintains the invariance with respect to the \chrtola{}
function.

Assume $a(\bar{X})\in\fchr(S)$.
If $a_r(\bar{X},\mathtt{p})\in\fchr(S)$ or $a_r(\bar{X},\mathtt{b})\in\fchr(S)$
then the following rule of $T(P)$ is applicable:
\[1 :: a_r(\bar{X},M)\ \backslash\ a(\bar{X})\iff M\neq\mathtt{n}\mid true\]
If $a_r(\bar{X},\mathtt{n})\in\fchr(S)$ then the rule below applies:
\[1 :: a_r(\bar{X},\mathtt{n}),a(\bar{X})\iff a_r(\bar{X},\mathtt{b})\]
Finally, if no 
rule of priority 1 can be applied, which implies that no constraint of the
form $a_r(\bar{X},M)\in\fchr(S)$, then the following $T(P)$ rule can fire:
\[2 :: a(\bar{X})\iff a_r(\bar{X},\mathtt{p})\]
This covers all possibilities. Now we look at what happens after firing one
of the priority 1 or 2 rules. The rule
\[1 :: a_r(\bar{X},M)\ \backslash\ a(\bar{X})\iff M\neq\mathtt{n}\mid\true\]
removes a constraint $a(\bar{X})\#i$ from $S$ and has an empty body,
so $\norm{\sigma'}=\norm{\sigma}-2$. Since $M\neq\mathtt{n}$ the removed 
constraint was already represented by the $a_r(\bar{X},M)$ constraint and so
$\chrtola(\sigma')=\chrtola(\sigma)$.
Firing
\[1 :: a_r(\bar{X},\mathtt{n}),a(\bar{X})\iff a_r(\bar{X},\mathtt{b})\]
causes the removal of two constraints from $S$, namely 
$a_r(\bar{X},\mathtt{n})\#i$ and $a(\bar{X})\#j$. Furthermore, it adds a new 
constraint $a_r(\bar{X},\mathtt{b})$ to $G$. This results in 
$\norm{\sigma'}=\norm{\sigma}-1$. The new constraint represents the combined 
mode of both removed constraints and hence 
$\chrtola(\sigma')=\chrtola(\sigma)$. 
Finally, the rule
\[2 :: a(\bar{X})\iff a_r(\bar{X},\mathtt{p})\]
is  only applicable if $\fchr(S)$ does not contain a constraint of the form
$a_r(\bar{X},M)$. It removes a constraint $a(\bar{X})\#i$ from $S$
and adds a new constraint $a_r(\bar{X},\mathtt{p})$ to $G$, resulting in
$\norm{\sigma'}=\norm{\sigma}-1$. The new representation covers the positive
assertion and so $\chrtola(\sigma')=\chrtola(\sigma)$.

In summary, if the goal is empty and $\sigma$ is not in pre-normal form, a
rule of priority $1$ or $2$ can fire and so no rule with lower priority is
applicable. All applicable transitions strictly decrease the value of the
ranking function and so the pre-normalization
terminates. Finally, none of the possible transitions changes the value of
\chrtola{}.\qed
\end{proof}
The state $\sigma^*$ is called a pre-normalization of $\sigma$.
\begin{definition}[Implied Rule Instance]
A rule instance $\theta(r)$ is implied in a state 
$\sigma$ if $\theta(C)\subseteq\chrtola(\sigma)$ with $\theta(\sigma)$ the
conclusion of $\theta(r)$.
\end{definition}

\begin{lemma}[Normalization]
Let there be given a pre-normalized state 
$\sigma=\langle \emptyset,S,\true,T\rangle_n$. If 
there exists a transition $\sigma\trans{\omegap}{T(P)}\sigma'$ in which an 
implied rule instance fires, then the pre-normalization of $\sigma'$ has the 
form $\langle\emptyset,S,\true,T'\rangle_{n'}$ with $T'\supsetneq T$. 
In other words $\chrtola(\sigma)=\chrtola(\sigma')$ and the CHR constraint
store after pre-normalization is unchanged from the one before the implied rule
instance fired while the propagation history is increased.
\label{lem:no-op}
\end{lemma}
\begin{proof}
Let $\theta(r)$ be the implied rule instance with conclusion $\theta(C)$.
Since $\theta(C)\subseteq\chrtola(\sigma)$ with 
$\sigma=\langle\emptyset,S,\true,T\rangle_n$, we have  
$\sigma'=\langle\theta(C),S,\true,T\cup\{t\}\rangle_n$ and 
$\chrtola(\sigma)=\chrtola(\sigma')$ with $t$ the propagation history tuple
corresponding to $\theta(r)$. The goal $G$ of $\sigma'$ equals $\theta(C)$ and
so it holds that if $a(\bar{X})\in G$ then $a(\bar{X},\mathtt{p})\in\fchr(S)$ 
or $a(\bar{X},\mathtt{b})\in\fchr(S)$ and if $del(a(\bar{X}))\in G$ then 
$a(\bar{X},\mathtt{n})\in\fchr(S)$ or $a(\bar{X},\mathtt{b})\in\fchr(S)$. Now
all constraints in the goal are first introduced in the CHR constraint store.
Next, the newly introduced CHR constraints are removed one by one using
one of the following rules:
\begin{align*}
1 :: a_r(\bar{X},M)\ \backslash\ a(\bar{X}) &\iff M \neq \mathtt{n}\mid\true\\
1 :: a_r(\bar{X},M)\ \backslash\ del(a(\bar{X}))&\iff M\neq\mathtt{p}\mid\true
\end{align*}
These rules remove all the constraints that were introduced from the goal and
do not change the rest of the CHR constraint store, hence after 
pre-normalization, the CHR constraint store equals that of state $\sigma$ 
again.\qed
\end{proof}
Because the CHR constraint store remains unchanged after firing an implied rule
instance and pre-normalizing the resulting state, only finitely many such
rule instances can fire before either reaching a final execution state, or a 
state in which a non-implied rule instance can fire. We call such a state
\emph{normalized}.
\begin{definition}[Normal Form]
A pre-normalized CHR execution state $\sigma$ is in normal form if it is a 
final state ($\sigma\notrans{\omegap}{T(P)}$) or there exists a
transition $\sigma\trans{\omegap}{T(P)}\sigma'$ such that
$\chrtola(\sigma')\nsupseteq\chrtola(\sigma)$, i.e., in which a non-implied 
rule instance is fired. 
\end{definition}
 
\begin{lemma}
For every Logical Algorithms state $\sigma_{LA}$ and every normalized 
CHR execution state $\sigma=\langle\emptyset,S,\true,T\rangle_n$ such that 
$\sigma_{LA}=\chrtola(\sigma)$, there exists a transition
$\sigma_{LA}\trans{LA}{P}\sigma'_{LA}$ if and only if there exists a transition
$\sigma\trans{\omegap}{T(P)}\sigma'$ firing a non-implied rule instance such
that $\sigma'_{LA}=\chrtola(\sigma')$.\label{lem:big-lem}
\end{lemma}
\begin{proof}
A transition of $\sigma_{LA}$ to $\sigma'_{LA}$ implies there exists a fireable
rule instance $\theta(r)$ of a rule $r$ in $P$ with priority $p$ of the form
\[r\ @\ p:A_1,\ldots,A_n\Rightarrow C\] 
Let $\langle A^u,A^c\rangle=
\langle [A^u_1,\ldots,A^u_m],[A^c_1,\ldots,A^c_l]\rangle
=\fsplit([A_1,\ldots,A_n])$ where
we use the \fsplit{} function defined in Section \ref{rule-trans}. 
The user-defined
antecedents can be partitioned into sets of syntactically equal antecedents
with respect to the matching substitution $\theta$. The following function
returns this partition:
\[\fsubtopar(\theta,[A^u_1,\ldots,A^u_m]) = \{S_1,\ldots, S_m\}\]
where $S_i=\{j\mid\theta(A^u_i)=\theta(A^u_j)\}$.
Let $\rho=\fsubtopar(\theta,A^u)$. From the 
partition, we find the most general unifier $\theta'$ that unifies all
antecedents $\{A^u_i\mid i\in S\}$ for every $S\in\rho$: 
$\theta'=\partmgu(\rho,A^u)$ with $\partmgu$ as defined in Section 
\ref{rule-trans}. Clearly, $\theta'$ exists and is more general than $\theta$.
The applicability of the \textbf{Apply} transition means that for all 
comparison antecedents $A_i^c$ with $1\leq i\leq l$, 
$\mathcal{D}\models\theta(A^c_i)$ and so it holds that 
$\mathcal{D}\models\bar{\exists}_\emptyset\theta'
(A^c_1\wedge\ldots\wedge A^c_l)$ and consequently a rule $r_\rho$ exists.
This rule looks as follows:
\[p+2 :: r_\rho\ @\ H_1,\ldots,H_k \implies g_1,g_2\mid C'\]
with $\langle [H_1,\ldots,H_k],g_1\rangle=\fmodes(A^f)$, 
$A^f=[A^f_1,\ldots,A^f_k]=\ffilter(\theta'(A^u),\rho)$, 
$g_2=\theta'(A^c)$ and $C'=\theta'(C)$. The \textsf{modes} and \textsf{filter}
functions are as defined in Section \ref{rule-trans}.

Let $\theta''$ be a ground matching substitution such that 
$\theta=\theta''|_{vars(\theta)}\circ\theta'$ where $\theta''|_{vars(\theta)}$
is the projection of $\theta''$ on the variables in $\theta$. Since $\theta'$
is more general than $\theta$, $\theta''$ exists. 
For all $i\in\{1,\ldots,k\}$, if $A^f_i=a(\bar{X})$ then
$H_i=a_r(\bar{X},\mathtt{p})$. Because of the applicability of Logical
Algorithms rule $r$ in state $\sigma_{LA}$, 
$\theta''(a(\bar{X}))\in \sigma_{LA}$ and $\theta''(del(a(\bar{X})))\notin
\sigma_{LA}$, so $H'_i=\theta''(a_r(\bar{X},\mathtt{p}))\#\textit{id}_i\in 
S$ and $\theta''(H_i)=\fchr(H'_i)$. Similarly, if $A^f_i=del(a(\bar{X}))$ then
$H_i=a_r(\bar{X},N)$ and $g_1$ contains $N\neq\mathtt{p}$; 
$\theta''(del(a(\bar{X})))\in \sigma_{LA}$ and as a result 
$H'_i=\theta''(a_r(\bar{X},N'))\#\textit{id}_i\in S$ with 
$N'=\mathtt{n}$ or $N'=\mathtt{b}$. Since $N$ only appears in $H_i$ and the
guard $N\neq\mathtt{p}$, we can further impose that $\theta''(N)=N'$ and then 
$\theta''(H_i)=\fchr(H'_i)$. 

All $\theta''(A^f_i)$ are different for $1\leq i\leq k$, and therefore, all 
$\textit{id}_i$ must be different. From 
$\mathcal{D}\models \bar{\exists}_\emptyset \theta(A^c_i)$ for $1\leq i\leq l$
and because $\theta''(g_1)=[N_1\neq\mathtt{p},\ldots,N_o\neq
\mathtt{p}]$ with $N_j=\mathtt{n}$ or $N_j=\mathtt{b}$ for $1\leq j\leq o$, 
$\mathcal{D}\models \true\rightarrow\bar{\exists}_\emptyset 
\theta''(g_1\wedge g_2)$. We conclude that $\theta''$ is a ground matching 
substitution that matches the head with constraints from $S$ and for which the
guard is entailed.

It is not possible that $\langle r_\rho,id(H)\rangle\in T$ because 
\chrtola{} grows monotonically, which implies that 
$\theta(C)=\theta''(C')\in \chrtola(\sigma)=\sigma_{LA}$
which contradicts with the applicability of $\theta(r)$ in $\sigma_{LA}$.

If we ignore rule priorities, all conditions are satisfied so that rule
instance $\theta(r_\rho)$ can fire. The resulting state $\sigma'$ has the form 
$\langle\theta(C),S,\true,T\cup\{\langle r_\rho,\fid(H)\rangle\}\rangle_n$. 
Clearly, if $\sigma_{LA}=
\chrtola(\langle\emptyset,S,true,T\rangle_n)$ and $\sigma'_{LA}=
\sigma_{LA}\cup\theta(C)$ then 
$\sigma'_{LA}=\chrtola(\sigma')$.
We now prove that every CHR transition firing a non-implied rule instance 
corresponds to a Logical Algorithms transition, also ignoring rule priorities.
Both results combined give us that the priority of the highest priority 
rule instance is equal in both $\sigma$ and $\sigma_{LA}$. 

A transition of $\sigma=\langle \emptyset,S,\true,T\rangle_n$ to $\sigma'$
implies that $T(P)$ contains a rule
\[p + 2 :: r_\rho\ @\ H\implies g_1,g_2\mid C'\]
and so the Logical Algorithms program $P$ contains a rule
\[r\ @\ p:A_1,\ldots,A_n\Rightarrow C\]
Let $\langle A^u,A^c\rangle=\fsplit([A_1,\ldots,A_n])$ and
$\theta=\partmgu(\rho,A^u)$. If $A_i=a(\bar{X})\in A^u$ then 
$\theta(a_r(\bar{X},\mathtt{p}))\in H$. If $A_i=del(a(\bar{X}))\in A^u$ then
$\theta(a_r(\bar{X},N))\in H$ and $(N\neq\mathtt{p})\in g_1$. Finally, if
$A_i\in A^c$ then $\theta(A_i)\in g_2$. There exists a (ground) matching 
substitution $\theta'$ such that $\theta'(H)\in\fchr(S)$ and 
$\mathcal{D}\models\bar{\exists}_\emptyset\theta'(g_1\wedge g_2)$. 

Let $\theta''=\theta'\circ\theta$ and let 
$\sigma_{LA}=\chrtola(\sigma)$.
Because $\theta'$ is a ground substitution, 
$\mathcal{D}\models\bar{\exists}_\emptyset\theta'(g_1\wedge g_2)$ implies that
for all $A_i\in A^c$, $\mathcal{D}\models\theta''(A_i)$. For all positive
user-defined antecedents $A_i=a(\bar{X})\in A^u$, we have that 
$\theta''(a(\bar{X},\mathtt{p}))\in\fchr(S)$ and so 
$\theta''(A_i)\in\sigma_{LA}$ and $del(\theta''(A_i))\notin\sigma_{LA}$. For
all negative user-defined antecedents $A_i=del(a(\bar{X}))\in A^u$, we have 
that $\theta''(a_r(\bar{X},N))\in\fchr(S)$ with $N=\mathtt{b}$ or $N=\mathtt{n}$
and so $\theta''(A_i)\in\sigma_{LA}$. We have assumed that $\theta'(r_\rho)$ is
not an implied rule instance and so 
$\theta'(C')=\theta''(C)\nsubseteq\sigma_{LA}$.

If we again ignore rule priorities, all conditions are satisfied so that rule
instance $\theta''(r)$ can fire in state $\sigma_{LA}$ and it holds that
$\sigma'_{LA}=\sigma_{LA}\cup\theta''(C)=\chrtola(\sigma')$ since
$\sigma'=\langle\theta'(C'),S,\true,T\cup\{\langle r_\rho,\fid(H)\rangle\}
\rangle_n$. Now we have that
both the original program $P$ and its translation $T(P)$ can fire corresponding
rule instances if we ignore priorities, and so their highest priority rule
instances also correspond.\qed
\end{proof}
\begin{theorem}
For every reachable \chrrp{} state $\sigma$, if 
$\sigma\trans{\omegap}{T(P)}\sigma'$ then either
$\chrtola(\sigma)=\chrtola(\sigma')$ or
$\chrtola(\sigma)\trans{LA}{P}\chrtola(\sigma')$.
\label{t:correspone}
\end{theorem}
\begin{proof}
Implied by Lemmas \ref{lem:prenorm}, \ref{lem:no-op} and \ref{lem:big-lem}.\qed
\end{proof}
\begin{theorem} For every Logical Algorithms state $\sigma_i$ and
reachable \chrrp{} state $\sigma'_i$ such that 
$\chrtola(\sigma'_i)=\sigma_i$, there
exists a finite \chrrp{} derivation
$\sigma'_i\deriv{\omegap}{T(P)}\sigma'_{i^*}$ for which holds that
$\chrtola(\sigma'_{i^*})=\sigma_i$ such that
if $\sigma_i\trans{LA}{P}\sigma_j$
then $\sigma'_{i^*}\trans{\omegap}{T(P)}\sigma'_j$
with $\chrtola(\sigma'_j)=\sigma_j$ and
if $\sigma_i$ is a final state then $\sigma'_{i^*}$ is also a final state.
\label{t:corresptwo}
\end{theorem}
\begin{proof}
Implied by Lemmas \ref{lem:prenorm}, \ref{lem:no-op} and \ref{lem:big-lem}.\qed
\end{proof}
Given a Logical Algorithms state $\sigma$, we can use 
$\langle \sigma,\emptyset,\true,\emptyset\rangle_1$ as initial state for the 
\chrrp{} derivation. 
Theorem \ref{t:corresptwo} is illustrated by the figure below.

\medskip{}
\begin{minipage}[l]{0.9\textwidth}
\begin{picture}(0,0)%
\includegraphics{correspondence.pstex}%
\end{picture}%
\setlength{\unitlength}{4144sp}%
\begingroup\makeatletter\ifx\SetFigFont\undefined%
\gdef\SetFigFont#1#2#3#4#5{%
  \reset@font\fontsize{#1}{#2pt}%
  \fontfamily{#3}\fontseries{#4}\fontshape{#5}%
  \selectfont}%
\fi\endgroup%
\begin{picture}(4343,1194)(1786,-1063)
\put(3376,-61){\makebox(0,0)[lb]{\smash{{\SetFigFont{10}{12.0}{\rmdefault}{\mddefault}{\updefault}
{\color[rgb]{0,0,0}$\sigma_i$}%
}}}}
\put(1801,-961){\makebox(0,0)[lb]{\smash{{\SetFigFont{10}{12.0}{\rmdefault}{\mddefault}{\updefault}
{\color[rgb]{0.375,0.375,0.375}$\langle\sigma_i,\emptyset,true,\emptyset\rangle_1
\deriv{\omega_p}{T(P)}$}{\color[rgb]{0,0,0}%
$\sigma'_i\deriv{\omega_p}{T(P)}\sigma'_{i^*}$}%
}}}}
\put(4591,-61){\makebox(0,0)[lb]{\smash{{\SetFigFont{10}{12.0}{\rmdefault}{\mddefault}{\updefault}
{\color[rgb]{0,0,0}$\sigma_i\trans{LA}{P}\sigma_j$}%
}}}}
\put(4501,-961){\makebox(0,0)[lb]{\smash{{\SetFigFont{10}{12.0}{\rmdefault}{\mddefault}{\updefault}
{\color[rgb]{0,0,0}$\sigma'_{i^*}\trans{\omega_p}{T(P)}\sigma'_j$}%
}}}}
\put(5761,-61){\makebox(0,0)[lb]{\smash{{\SetFigFont{10}{12.0}{\rmdefault}{\mddefault}{\updefault}
{\color[rgb]{0,0,0}$\sigma_i\notrans{LA}{P}$}%
}}}}
\put(5716,-961){\makebox(0,0)[lb]{\smash{{\SetFigFont{10}{12.0}{\rmdefault}{\mddefault}{\updefault}
{\color[rgb]{0,0,0}$\sigma'_{i^*}\notrans{\omega_p}{T(P)}$}%
}}}}
\end{picture}%

\end{minipage}
\subsection{Weak Bisimulation}
To capture the meaning of the above correspondence results, we relate them to
the notion of (weak) bisimulation. A bisimulation is a relation between the
states of a labeled transition system (LTS). A relation 
$R\subseteq S_1\times S_2$
between the states in $S_1$ and those in $S_2$ is a bisimulation if 
$p\ R\ q$ and $p\ltstrans{\alpha} p'$ implies that 
$q\ltstrans{\alpha} q'$ with $p'\ R\ q'$, and similarly,
$p\ R\ q$ and $q\ltstrans{\alpha} q'$ implies that 
$p\ltstrans{\alpha} p'$ with $p'\ R\ q'$. Here, $\alpha$ is the
label of the transition $p\ltstrans{\alpha} p'$ from state $p$ to
state $p'$. If a transition from $p$ to $p'$ has no observable effect, it is
called a \emph{silent} transition and denoted by 
$p\ltstrans{\tau} p'$. A relation $R\subseteq S_1\times S_2$
is a \emph{weak} bisimulation if $p\ R\ q$ and 
$p\ltstrans{\alpha} p'$ implies that 
$q\ltstrans{\tau}^* q_*\ltstrans{\alpha} q'_* \ltstrans{\tau}^* q'$ with 
$p'\ R\ q'$, and vice versa with the
roles of $p$ and $q$ swapped. Here $p\ltstrans{\tau}^*p'$ means
$p$ and $p'$ are linked by zero or more silent transitions.

Let $S_1$ be the set of valid Logical Algorithms states for program $P$ and let 
$S_2=\{\chrtola(\sigma)\mid 
\langle G,\emptyset,true,\emptyset\rangle_1\deriv{\omegap}{T(P)}\sigma\wedge
G\in S_1\}$, i.e., $S_2$ is found by applying the \chrtola{} mapping function
all reachable \chrrp{} states for program $T(P)$. 
We transform the state transition systems for Logical 
Algorithms and \chrrp\ to labeled transition systems as follows: a 
Logical Algorithms transition
$\sigma \trans{LA}{P}\sigma'$ corresponds to an LTS transition 
$\sigma \ltstrans{\alpha}\sigma'$ with 
$\alpha=\sigma'\setminus\sigma$, i.e., $\alpha$ represents the state change
from $\sigma$ to
$\sigma'$. A \chrrp\ transition 
$\sigma\trans{\omegap}{T(P)}\sigma'$ corresponds to an LTS transition
$\chrtola(\sigma)\ltstrans{\alpha}\chrtola(\sigma')$ with
$\alpha=\chrtola(\sigma')\setminus\chrtola(\sigma)$ if this set is not empty
and $\alpha=\tau$ otherwise.
\begin{corollary}
The equality relation between the states of $S_1$ and $S_2$ is a weak 
bisimulation.
\end{corollary}

\section{Translating a subset of \chrrp\ into Logical Algorithms\label{chr-to-la}}
In the previous section, we have shown that Logical Algorithms programs can be 
translated into equivalent \chrrp{} programs. In this section, we show how to 
do the opposite, i.e., how  \chrrp{} programs can be translated into equivalent
Logical Algorithms programs. This allows us to apply the meta-complexity
theorem for Logical Algorithms to the translation of these \chrrp{} programs.

We impose some restrictions on the \chrrp{} programs that can be translated.
These restrictions result from the fact that the Logical Algorithms
language does not have the concept of an underlying constraint solver that 
offers both ask and tell built-in constraints. In principle, the complete
\chrrp{} language could be translated into LA, as the subset of \chrrp{} for
which we propose a translation is already Turing complete 
\cite[Chapter~10]{jon:thesis}, and therefore so is the LA language. However, in
general, this requires a LA implementation of the built-in constraint solver
used by the \chrrp{} program. Since this built-in solver is not part of the 
\chrrp{} program, we restrict our translation schema to programs that do not 
make use of such a solver. In particular, we only support
the translation of the \emph{positive range-restricted ground} segment of
\chrrp{} \cite{hariolf:petri}:
\begin{enumerate}
\item In all \emph{reachable} states $\sigma=\langle G,S,B,T\rangle_n$:
$vars(S)=\emptyset$. In words, all (stored) CHR constraints are ground.
\item All built-in constraints are comparisons; there are no built-in tell
constraints.
\end{enumerate}
The first property holds if the initial goal is ground and all rules are 
\emph{variable restricted}, which means that all variables in the body of a 
rule, also appear in one of the rule heads. The second property implies that
all reachable states are of the form $\langle G,S,\true,T\rangle_n$, i.e., the
built-in constraint store always equals \true.

To simplify the presentation, we also assume that the priority of dynamic
priority rules is determined by the arguments of its left-most head. In 
general, we can use the transformation schema given in 
Section~\ref{sec:la_and_chr:chr:differences} to ensure that the resulting 
Logical Algorithms rules have the correct syntactical form. 
\subsection{The Translation Schema}
We now show how the rules of a \chrrp{} program $P$ are transformed into 
Logical Algorithms rules that form a program $T(P)$.
To increase readability, we distinguish between simplification and simpagation
rules on the one hand, and propagation rules on the other. A \emph{simpagation}
rule of the form
\[p :: r\ @\ H_1,\ldots,H_{l-1}\backslash H_l,\ldots,H_m \iff 
g\mid B_1,\ldots,B_o\]
is transformed into
\begin{align*}
r'\ @\ p:H_1^\textit{id},\ldots,H_m^\textit{id},
    \textit{All}&\textit{diff},g,\textit{next\_id}(\textit{Id}_\textit{next})
 \Rightarrow\\
&\textit{del}(H_l^\textit{id}),\ldots,\textit{del}(H_m^\textit{id}),
\textit{del}(\textit{next\_id}(\textit{Id}_\textit{next})),\\
&B^\textit{id}_1,\ldots,B^\textit{id}_o, 
\textit{next\_id}(\textit{Id}_\textit{next}+o)
\end{align*}
where $H_i^{id}=\texttt{c($\bar{X}$,$Id_i$)}$
if $H_i=\texttt{c($\bar{X}$)}$, 
$B^{id}_i=\texttt{c($\bar{X}$,$Id_{next}+i-1$)}$ if $B_i=\texttt{c($\bar{X}$)}$
and $\textit{Alldiff}=\{(Id_i\neq Id_j)\mid\mathcal{D}\models
\bar{\exists}_\emptyset H_i=H_j\wedge g\}$. The disequalities
in \textit{Alldiff} are between those heads that are unifiable and for which 
the guard is still satisfiable after this unification. The \verb!next_id!/1
antecedent is used to retrieve the next free identifier to be used to identify
constraints in the body, cf.\ the \omegap{} operational semantics of \chrrp{}.
The case of a 
simplification rule is similar. A propagation rule of the form
\[p :: r\ @\ H_1,\ldots,H_m \implies g\mid B_1,\ldots,B_o\]
is transformed into the following two rules
\begin{align*}
r'_1\ @\ p:H_1^\textit{id},\ldots,H_m^\textit{id},
    \textit{All}&\textit{diff},g \Rightarrow 
    \textit{token}(r,[\textit{Id}_1,\ldots,\textit{Id}_m])\\
r'_2\ @\ p:H_1^\textit{id},\ldots,H_m^\textit{id}, 
    \textit{All}&\textit{diff},g, 
    \textit{token}(r,[\textit{Id}_1,\ldots,\textit{Id}_m]), 
    \textit{next\_id}(\textit{Id}_\textit{next})
 \Rightarrow\\
&\textit{del}(\textit{token}(r,[\textit{Id}_1,\ldots,\textit{Id}_m])),
\textit{del}(\textit{next\_id}(\textit{Id}_\textit{next})),\\
&B^\textit{id}_1,\ldots,B^\textit{id}_o, 
\textit{next\_id}(\textit{Id}_\textit{next}+o)
\end{align*}
where $H_i^{id}$, $B_i^{id}$ and \textit{Alldiff} are as before. The first of
these rules generates a token. This token is removed by the second rule. The
tokens are needed to prevent a given rule instance from firing more than once.%
\footnote{In \cite{chrla}, an erroneous translation was presented which did not
use tokens, and in which a propagation rule could fire infinitely many times 
because the constraints in the body are assigned new identifiers each time the
rule is fired.} Note that the transformation into two rules and the use of
tokens does not increase the complexity compared to the original rule, as there
is only one token for each combination of rule and constraint identifiers (as 
well as only one positive \verb!next_id!/1 assertion in any state). 

The initial database consists of the goal (where each constraint is extended
with a unique identifier) and a \verb!next_id(!$Id_{next}$\verb!)! assertion
(with $Id_{next}$ the next free identifier).
\begin{example}[Merge Sort]\label{ex:chrrp_to_la:merge_sort}
The following \chrrp{} program implements a merge sort algorithm. Its input
consists of a series of $n$ (a power of 2) \verb!number!/1 constraints.
Its output is a sorted
list of the numbers in the input, represented as \verb!arrow!/2 constraints,
where \texttt{arrow($X$,$Y$)} indicates that $X$ is right before
$Y$.
\begin{Verbatim}[frame=single,fontsize=\small]
1 :: ms1 @ arrow(X,A) \ arrow(X,B) <=> A < B | arrow(A,B).
2 :: ms2 @ merge(N,A),  merge(N,B) <=> A < B | merge(2*N+1,A), arrow(A,B).
3 :: ms3 @ number(X) <=> merge(0,X).
\end{Verbatim}
Its Logical Algorithms translation is
\begin{Verbatim}[frame=single,fontsize=\small,commandchars=\\\{\}]
ms1\vrbprime @ 1 : arrow(X,A,Id\vrbsubt{1}), arrow(X,B,Id\vrbsubt{2}), A < B, next_id(NId) =>
            del(arrow(X,B,Id\vrbsubt{2})), del(next_id(NId)), 
            arrow(A,B,NId), next_id(NId+1).
ms2\vrbprime @ 2 : merge(N,A,Id\vrbsubt{1}), merge(N,B,Id\vrbsubt{2}), A < B, next_id(NId) =>
            del(merge(N,A,Id\vrbsubt{1})), del(merge(N,B,Id\vrbsubt{2})), del(next_id(NId)),
            merge(2*N+1,A,NId), arrow(A,B,NId+1), next_id(NId+2).
ms3\vrbprime @ 3 : number(X,Id), next_id(NId) => del(number(X,Id)), 
            del(next_id(NId)), merge(0,X,NId), next_id(NId+1).
\end{Verbatim}
Note that in rules \verb!ms1! and \verb!ms2!, the guard prevents the 
constraints matching the heads from being equal, and so there
are no disequality constraints between the CHR constraint identifiers.
In \cite{chrla} it is derived that the total runtime of this Logical Algorithms
program is $\mathcal{O}(n\log n)$. We defer the complexity analysis of the
merge sort algorithm to Section~\ref{sub:new_meta:examples} where we analyse 
the \chrrp{} implementation directly using a new meta-complexity theorem for
\chrrp{}.
\end{example}
\begin{example}[Less-or-Equal]
To illustrate how propagation rules are dealt with, we show the translation
of a rule of the \texttt{leq} program which is given further on in 
Example~\ref{ex:new_meta:leq}. The rule
\begin{Verbatim}[frame=single,fontsize=\small]
3 :: transitivity @ leq(X,Y), leq(Y,Z) ==> leq(X,Z).
\end{Verbatim}
is translated into
\begin{Verbatim}[frame=single,fontsize=\small,commandchars=*\{\}]
transitivity*vrbprimesubt{1} @ 3 : leq(X,Y,Id*vrbsubt{1}), leq(Y,Z,Id*vrbsubt{2}), Id*vrbsubt{1} \= Id*vrbsubt{2} =>
                token(transitivity,[Id*vrbsubt{1},Id*vrbsubt{2}]).
transitivity*vrbprimesubt{2} @ 3 : leq(X,Y,Id*vrbsubt{1}), leq(Y,Z,Id*vrbsubt{2}), Id*vrbsubt{1} \= Id*vrbsubt{2},
        token(transitivity,[Id*vrbsubt{1},Id*vrbsubt{2}]), next_id(NId) =>
                del(token(transitivity,[Id*vrbsubt{1},Id*vrbsubt{2}])), del(next_id(NId)),
                leq(X,Z,NId), next_id(NId+1).
\end{Verbatim}
Note that since in the original rule, the two heads \verb!leq(X,Y)! and 
\verb!leq(Y,Z)! are unifiable (and there is furthermore no guard to prevent 
this from happening), we have to add an explicit disequality between the 
constraint identifiers for these heads: 
\verb!Id!\vrbsubt{1}\verb! \= Id!\vrbsubt{2}.
\end{example}

\subsection{Correspondence}
In this subsection, we prove that a \chrrp{} program and its translation to 
Logical Algorithms are operationally equivalent. Again we introduce a mapping
function:
\[\latochr(\sigma)=\langle\emptyset,S,\true,T\rangle_n\]
where the CHR constraint store $S=\{c(\bar{X})\#\varid\mid c(\bar{X},\varid)\in
\sigma\wedge\del(c(\bar{X},\varid))\notin\sigma\}$, the propagation history 
$T=\{\langle R,\varids\rangle\mid\del(\token(R,\varids))\in\sigma\}$, and the
next free identifier $n$ is such that $\textit{next\_id}(n)\in\sigma$ and
$\del(\textit{next\_id}(n))\notin\sigma$.
In the following, we consider a Logical Algorithms state $\sigma$ reachable
with respect to program $T(P)$
if it can be derived from an initial state consisting of CHR constraints 
extended with unique identifiers, and a single \verb!next_id!/1 assertion with
as argument the next free identifier. In this case, reachability amongst others
implies that there can be only one (strictly) positive \verb!next_id!/1 
assertion in any state, and no two CHR constraint representations share their
identifier.

First, we define the \emph{priority-ignoring} operational semantics 
$LA'$ of Logical Algorithms, as being the same as its regular operational
semantics except that priorities are ignored, i.e., the next applicable rule
instance in any state is independent of the priorities. Next, we state two
lemmas that relate a \chrrp{} program $P$ and its translation $T(P)$ under
respectively the theoretical operational semantics \omegat{} of CHR and this
priority-ignoring operational semantics $LA'$ of Logical Algorithms. Note
that the \omegat{} semantics of CHR also ignores rule priorities.
\begin{lemma}\label{lem:chrrp_to_la:la_implies_chrrp}
For every reachable Logical Algorithms state $\sigma_i$ it holds that if 
$\sigma_i\trans{LA'}{T(P)}\sigma_j$, then either it holds that 
$\latochr(\sigma_i)=\latochr(\sigma_j)$ or there exists a finite CHR derivation
$\latochr(\sigma_i)=\langle\emptyset,S,\true,T\rangle_n\trans{\omegat}{P}
\langle C,S',\true,T'\rangle_n\deriv{\omegat}{P}\langle\emptyset,S'',\true,
T'\rangle_{n'}=\latochr(\sigma_j)$
consisting of an \textbf{Apply} transition,
followed by zero or more \textbf{Introduce} transitions.
\end{lemma}
\begin{proof}
Consider a transition $\sigma_i\trans{LA'}{T(P)}\sigma_j$. The only type of 
transition in Logical Algorithms is the \textbf{Apply} transition which fires
a rule. If $\latochr(\sigma_i)=\latochr(\sigma_j)$, then this rule must be of
the form
\[r'_1\ @\ p:H_1^\textit{id},\ldots,H_m^\textit{id},\textit{Alldiff},g 
    \Rightarrow\textit{token}(r,[\textit{Id}_1,\ldots,\textit{Id}_m])\]
because all other types of rules either delete the representation of a CHR
constraint which changes the CHR constraint store, or remove a token which
results in an extended propagation history. We call the fired rule a
\emph{token generation rule}.

If $\latochr(\sigma_i)\neq
\latochr(\sigma_j)$ and the rule fired is of the form
\begin{align*}
r'\ @\ p:H_1^\textit{id},\ldots,H_m^\textit{id},
    \textit{All}&\textit{diff},g,\textit{next\_id}(\textit{Id}_\textit{next})
 \Rightarrow\\
&\textit{del}(H_l^\textit{id}),\ldots,\textit{del}(H_m^\textit{id}),
\textit{del}(\textit{next\_id}(\textit{Id}_\textit{next})),\\
&B^\textit{id}_1,\ldots,B^\textit{id}_o, 
\textit{next\_id}(\textit{Id}_\textit{next}+o)
\end{align*}
which corresponds to a simplification ($l=1$) or simpagation ($l>1$) rule. We
further assume the case of a simpagation rule; the case of a simplification
rule is similar. If $r'\in T(P)$ (with $l>1$), then $P$ contains a rule
\[p::r\ @\ H_1,\ldots,H_{l-1}\backslash H_l,\ldots,H_m\iff
    g\mid B_1,\ldots,B_o\]
Since the conditions for the Logical Algorithms \textbf{Apply} transition are
satisfied, there exists a ground matching substitution $\theta$ such that for
each antecedent $H_i^\mathit{id}=c(\bar{X},\varid_i)$ ($1\leq i\leq m$) 
it holds that
$\theta(H_i^\mathit{id})\in\sigma$ and $\del(\theta(H_i^\mathit{id}))
\notin\sigma$ and so
by definition of the \latochr{} function, $\theta(H_i\#\varid_i)\in S$
where
$\latochr(\sigma_i)=\sigma'_i=\langle\emptyset,S,\true,T\rangle_n$. For each
comparison $g_i\in g$, it holds that $\mathcal{D}\models\theta(g_i)$ and so
$\mathcal{D}\models\true\rightarrow\bar{\exists}_\emptyset
\theta(g)$. Since $r$ is a simpagation rule, the propagation history $T$ does
not contain any element of the form $\langle r,\avar\rangle$. In summary, 
all conditions are satisfied such that the rule instance $\theta(r)$ can fire
in state $\sigma'_i$ under operational semantics \omegat{}. 

After firing $\theta(r)$ in state $\sigma'_i$, the resulting state equals
$\langle\theta(B_1\wedge\ldots\wedge B_o),S',\true,T\rangle_n$ where 
$S'=S\setminus\{\theta(H_l\#\varid_l),\ldots,\theta(H_m\#\varid_m)\}$.
In this state, we can apply the \textbf{Introduce} $o$ times before reaching a
state with an empty 
goal. There are $o!$ 
possible orders in which the introductions can be applied; the one we need
is the order in which the $B_i$ constraints appear in the rule body. Following
this order, the state resulting from the introductions equals $\sigma'_j=
\langle\emptyset,S'',\true,T\rangle_{(n+o)}$ where $S''=S'\cup\{\theta(B_1)\#
n,\ldots,\theta(B_o)\# (n+o-1)\}$. 
It is easy to see that this state $\sigma'_j$ 
equals $\latochr(\sigma_j)$, the state resulting from firing Logical Algorithms
rule instance $\theta(r')$ in state $\sigma_i$.

If $\latochr(\sigma_i)\neq\latochr(\sigma_j)$ and the rule fired is not of the
form shown above, then it must have the following form
\begin{align*}
r'_2\ @\ p:H_1^\textit{id},\ldots,H_m^\textit{id}, 
    \textit{All}&\textit{diff},g, 
    \textit{token}(r,[\textit{Id}_1,\ldots,\textit{Id}_m]), 
    \textit{next\_id}(\textit{Id}_\textit{next})
 \Rightarrow\\
&\textit{del}(\textit{token}(r,[\textit{Id}_1,\ldots,\textit{Id}_m])),
\textit{del}(\textit{next\_id}(\textit{Id}_\textit{next})),\\
&B^\textit{id}_1,\ldots,B^\textit{id}_o, 
\textit{next\_id}(\textit{Id}_\textit{next}+o)
\end{align*}
the corresponding \chrrp{} rule in $P$ looks like
\[p :: r\ @\ H_1,\ldots,H_m \implies g\mid B_1,\ldots,B_o\]
Again, since the conditions for the Logical Algorithms \textbf{Apply} 
transition are satisfied, there exists a ground matching substitution $\theta$
such that for each antecedent $H_i^\mathit{id}=c(\bar{X},\varid_i)$ 
($1\leq i\leq m$) 
in rule $r'_2$ it holds that $\theta(H_i^\mathit{id})\in\sigma$ and 
$\del(\theta(H_i^\mathit{id}))\notin\sigma$ and so by definition of the
\latochr{} function, $\theta(H_i\#\varid_i)\in S$ where
$\latochr(\sigma_i)=\sigma'_i=\langle\emptyset,S,\true,T\rangle_n$. For each
comparison $g_i\in g$, it holds that $\mathcal{D}\models\theta(g_i)$ and so 
$\mathcal{D}\models\true\rightarrow\bar{\exists}_\emptyset
\theta(g)$. The propagation history $T$ cannot contain 
$\langle r,\theta([\varid_1,\ldots,\varid_m])\rangle$ because by definition of
the \latochr{} function this would imply that the atom 
$\token(r,\theta([\varid_1,\ldots,\varid_m))$ was deleted in some earlier 
state, which contradicts with the applicability of the \textbf{Apply} 
transition on
rule instance $\theta(r'_2)$. Again, all
conditions are satisfied such that $\theta(r)$ can fire in state $\sigma'_i$.

After firing $\theta(r)$ in state $\sigma'_i$, the resulting state equals
$\langle\theta(B_1\wedge \ldots\wedge B_o),S,\true,T'\rangle_n$ where 
$T'=T\cup\{\langle r,[\varid_1,\ldots,\varid_m]\rangle\}$. 
In this state, we can apply the \textbf{Introduce} transition $o$ times before
reaching a state with an empty 
goal. Given again that these introductions are applied in the order in which
the $B_i$ constraints appear in the rule body, then the resulting state equals
$\sigma'_j=\langle\emptyset,S',\true,T'\rangle_{(n+o)}$ where 
$S'=S\cup\{\theta(B_1)\# n,\ldots,\theta(B_o)\# (n+o-1)\}$. It is again easy to
see 
that this state $\sigma'_j$ equals $\latochr(\sigma_j)$, the state resulting 
from firing Logical Algorithms rule instance $\theta(r'_2)$ in state 
$\sigma_i$.\qed 
\end{proof}
\begin{lemma}\label{lem:chrrp_to_la:chrrp_implies_la}
For every reachable \chrrp{} state $\sigma_i$ and reachable Logical Algorithms
state $\sigma'_i$ with $\latochr(\sigma'_i)=\sigma_i$, there exists a finite 
Logical Algorithms derivation $\sigma'_i\deriv{LA'}{T(P)}\sigma'_{i^*}$ with
$\latochr(\sigma'_{i^*})=\sigma_i$ such that if 
$\sigma_i=\langle\emptyset,S,\true,T\rangle_n\trans{\omegat}{P}
\langle C,S',\true,T'\rangle_n\deriv{\omegat}{P}$ $
\langle \emptyset,S'',\true,T'\rangle_{n'}=\sigma_j$ where the derivation 
consists of a single \textbf{Apply} transition, followed by zero or more 
\textbf{Introduce} transitions, then $\sigma'_{i^*}\trans{LA'}{T(P)}\sigma'_j$
with $\latochr(\sigma'_j)=\sigma_j$ and if $\sigma_i$ is a final state then 
$\sigma'_{i^*}$ is also a final state.
\end{lemma}
\begin{proof}
Let there be given a reachable Logical Algorithms state $\sigma'_i$ with 
$\latochr(\sigma'_i)=\sigma_i$. Because of 
Lemma~\ref{lem:chrrp_to_la:la_implies_chrrp}, state $\sigma_i$ is also
reachable in \chrrp{} with respect to program $P$. Assume
$\sigma_i=\langle\emptyset,S,\true,T\rangle_n\trans{\omegat}{P}
\langle C,S',\true,T'\rangle_n\deriv{\omegat}{P}
\langle \emptyset,S'',\true,T'\rangle_{n'}=\sigma_j$ where the derivation 
consists of a single \textbf{Apply} transition, followed by zero or more 
\textbf{Introduce} transitions, and let $\theta(r)$ be the \chrrp{} rule 
instance that fired in state $\sigma_i$. If $r$ is simplification ($l=1$) or
simpagation ($l>1$) rule
\[p :: r\ @\ H_1,\ldots,H_{l-1}\backslash H_l,\ldots,H_m\iff g\mid B_1,\ldots,
    B_o\]
then $\theta(H_i)\#\textit{id}_i\in S$ for $1\leq i\leq m$ with 
$\textit{id}_i\neq \textit{id}_j$ if $i\neq j$, and
$\mathcal{D}\models \bar{\exists}_\emptyset \theta(g)$. Furthermore, $T(P)$ 
contains a rule
\begin{align*}
r'\ @\ p:H_1^\textit{id},\ldots,H_m^\textit{id},
    \textit{All}&\textit{diff},g,\textit{next\_id}(\textit{Id}_\textit{next})
 \Rightarrow\\
&\textit{del}(H_l^\textit{id}),\ldots,\textit{del}(H_m^\textit{id}),
\textit{del}(\textit{next\_id}(\textit{Id}_\textit{next})),\\
&B^\textit{id}_1,\ldots,B^\textit{id}_o, 
\textit{next\_id}(\textit{Id}_\textit{next}+o)
\end{align*}
Now let $\theta'$ be a ground matching substitution such that 
$\theta'|_{vars(\theta)}=\theta$ where $\theta'_{vars(\theta)}$ is the 
projection of $\theta'$ on the variables in $\theta$, and such that both
$\theta'(\textit{Id}_i)=\textit{id}_i$ for $1\leq i\leq m$ and
$\theta'(\textit{Id}_\textit{next})=n$. Since for $1\leq i\leq m$, 
$H_i^\mathit{id}=c(\bar{X},\mathit{Id}_i)$ if $H_i=c(\bar{X})$, it holds that
$\theta'(H_i^\mathit{id})\in\sigma'_i$ and
$del(\theta'(H_i^\mathit{id}))\notin\sigma'_i$. Also, 
$\mathcal{D}\models\bar{\exists}_\emptyset\theta(g)$ implies
$\mathcal{D}\models \theta(g_i)$ for each comparison $g_i\in g$.
Note that because $\theta(g)$ is ground, there is no existential
quantification. The \textit{Alldiff} conditions hold because 
$\theta'(\mathit{Id}_i)=\theta'(\mathit{Id}_j)$ implies that $i=j$. Finally,
because of the reachability of state $\sigma'_i$, there is exactly one
strictly positive \verb!next_id!/1 assertion in $\sigma'_i$ whose argument
equals $n$. Finally, the rule conclusion cannot be already included in the
state $\sigma'_i$ because it includes amongst others the deletion of at least
one of the antecedents. Therefore, all conditions are satisfied
such that rule instance $\theta'(r')$ can fire in state $\sigma'_i$, resulting
in a state $\sigma'_j=\latochr(\sigma_j)$. 

Now assume that in the \chrrp{} state $\sigma_i$, a rule instance $\theta(r)$
fires where $r$ is a propagation rule:
\[p :: r\ @\ H_1,\ldots,H_m\implies g\mid B_1,\ldots, B_o\]
In this case the Logical Algorithms translation $T(P)$ contains the following
rules:
\begin{align*}
r'_1\ @\ p:H_1^\textit{id},\ldots,H_n^\textit{id},
    \textit{All}&\textit{diff},g \Rightarrow 
    \textit{token}(r,[\textit{Id}_1,\ldots,\textit{Id}_n])\\
r'_2\ @\ p:H_1^\textit{id},\ldots,H_n^\textit{id}, 
    \textit{All}&\textit{diff},g, 
    \textit{token}(r,[\textit{Id}_1,\ldots,\textit{Id}_n]), 
    \textit{next\_id}(\textit{Id}_\textit{next})
 \Rightarrow\\
&\textit{del}(\textit{token}(r,[\textit{Id}_1,\ldots,\textit{Id}_n])),
\textit{del}(\textit{next\_id}(\textit{Id}_\textit{next})),\\
&B^\textit{id}_1,\ldots,B^\textit{id}_l, 
\textit{next\_id}(\textit{Id}_\textit{next}+l)
\end{align*}
A similar analysis as above shows that there exists a matching substitution
$\theta'$ with $\theta'|_{vars(\theta)}=\theta$ and both
$\theta'(\textit{Id}_i)=\textit{id}_i$ for $1\leq i\leq m$ and
$\theta'(\textit{Id}_\textit{next})=n$, such that rule instance 
$\theta'(r'_1)$ can fire (ignoring priorities) if 
$\mathit{token}(r,[\mathit{id}_1,\ldots,\mathit{id}_n])\notin \sigma'_i$
and $\theta'(r'_2)$ otherwise. If $\theta'(r'_1)$ fires then the resulting 
state $\sigma'_{i^*}=\sigma'_i\cup 
\{\mathit{token}(r,[\mathit{id}_1,\ldots,\mathit{id}_n])\}$ and clearly
$\latochr(\sigma'_{i^*})=\latochr(\sigma'_i)$. Moreover, in state 
$\sigma'_{i^*}$, rule instance $\theta'(r'_2)$ can fire and for the resulting
state $\sigma'_j$ it holds that $\latochr(\sigma'_j)=\sigma_j$. If already
$\mathit{token}(r,[\mathit{id}_1,\ldots,\mathit{id}_n])\in \sigma'_i$ then the
same reasoning holds with $\sigma'_i=\sigma'_{i^*}$.

Finally, assume that \chrrp{} state $\sigma_i$ is a final state. If $\sigma'_i$
is not a final Logical Algorithms state, then because of 
Lemma~\ref{lem:chrrp_to_la:la_implies_chrrp}, the only applicable rules are
those that do not change the result of the $\latochr$ function. Only the
token generation rules satisfy this property. Since they only generate tokens
and these tokens do not appear in their antecedents, these rules can fire only
finitely many times before a final Logical Algorithms state $\sigma_{i^*}$ is
reached.
\end{proof}
Finally, we state two theorems that essentially are the same as the lemmas
above, except that they do take into account rule priorities.
\begin{theorem}\label{the:chrrp_to_la:la_implies_chrrp}
For every reachable Logical Algorithms state $\sigma_i$ it holds that if 
$\sigma_i\trans{LA}{T(P)}\sigma_j$, then either it holds that 
$\latochr(\sigma_i)=\latochr(\sigma_j)$ or there exists a finite CHR derivation
$\latochr(\sigma_i)=\langle\emptyset,S,\true,T\rangle_n\trans{\omegap}{P}
\langle C,S',\true,T'\rangle_n\deriv{\omegap}{P}\langle\emptyset,S'',\true,
T'\rangle_{n'}=\latochr(\sigma_j)$
consisting of an \textbf{Apply} transition,
followed by zero or more \textbf{Introduce} transitions.
\end{theorem}
\begin{theorem}\label{the:chrrp_to_la:chrrp_implies_la}
For every reachable \chrrp{} state $\sigma_i$ and reachable Logical Algorithms
state $\sigma'_i$ with $\latochr(\sigma'_i)=\sigma_i$, there exists a finite 
Logical Algorithms derivation $\sigma'_i\deriv{LA}{T(P)}\sigma'_{i^*}$ with
$\latochr(\sigma'_{i^*})=\sigma_i$ such that if 
$\sigma_i=\langle\emptyset,S,\true,T\rangle_n\trans{\omegap}{P}
\langle C,S',\true,T'\rangle_n\deriv{\omegap}{P}$ $
\langle \emptyset,S'',\true,T'\rangle_{n'}=\sigma_j$ where the derivation 
consists of a single \textbf{Apply} transition, followed by zero or more 
\textbf{Introduce} transitions, then $\sigma'_{i^*}\trans{LA}{T(P)}\sigma'_j$
with $\latochr(\sigma'_j)=\sigma_j$ and if $\sigma_i$ is a final state then 
$\sigma'_{i^*}$ is also a final state.
\end{theorem}
\begin{proof}
Both theorems are implied by Lemmas \ref{lem:chrrp_to_la:la_implies_chrrp} and
\ref{lem:chrrp_to_la:chrrp_implies_la}, combined with the fact that in 
corresponding states, corresponding rules can fire which have the same 
priority. Therefore, the highest priority applicable rule instances are also
equal in corresponding states.
\end{proof}
\section{Implementing \chrrp, the Logical Algorithms way}
\label{sec:new_implementation}
This section presents a new implementation for \chrrp{}, based on the 
implementation proposal for Logical Algorithms presented in \cite{la}, as well
as on the scheduling algorithm presented in \cite{mergeable_schedules}. The
purpose of this implementation is not to replace our existing \chrrp{} 
implementation as presented in \cite{compiling_chrrp}, but to 
support a new meta-complexity theorem for \chrrp{}, based on the result for 
Logical Algorithms, and extended towards the full \chrrp{} language. This
includes in particular support for non-ground constraints and a built-in
constraint theory. We note that a better worst case complexity 
for certain operations is not always worthwhile in practice due to larger 
constant factors in the average case. Also, the proposed implementation may not
always achieve a better complexity than the existing implementation. The main
purpose remains to have a relatively straightforward way to derive for a given
\chrrp{} program, a bound that is guaranteed to be an upper bound for at least
the implementation proposed. Since the meta-complexity result is insensitive to
constant factors, we can present the new implementation as a source-to-source
transformation to regular CHR.

The proposed implementation consists of the compilation of the \chrrp{} rules
of the input program into regular CHR rules in which matching is made explicit,
combined with a scheduler module that is responsible for the execution control.
The implementation is correct if it is executed according to the refined
operational semantics of CHR \cite{duck:refined}, which describes the execution
strategy
followed by most current CHR implementations. We have based our implementation
on the high-level implementation proposal for Logical Algorithm of \cite{la},
extended where necessary to support general built-in constraints. By
using a CHR implementation with advanced indexing support, like for example the
K.U.Leuven CHR system \cite{kulchr}, our implementation also offers strong
complexity guarantees that facilitate a new meta-complexity theorem for 
\chrrp{}, similar to the one for Logical Algorithms (see 
Section~\ref{sec:new_meta}). In the following, we make use of Prolog as 
CHR's host language, but the implementation can easily be adapted to work with
a different host language.
\subsection{Overview}\label{sec:new_implementation:overview}
The implementation is based on a form of lazy (on-demand) matching with 
retainment of previously computed partial matches. It combines the concept of
alpha and beta memories from the RETE algorithm \cite{rete}, with lazy matching
as for example implemented by the LEAPS algorithm \cite{leaps}.%
\footnote{Most current CHR systems, including the K.U.Leuven CHR system and
the \chrrp{} system of \cite{compiling_chrrp}, use a variant of the LEAPS
algorithm for rule matching.} The basic idea is as follows. A new constraint 
can function both as a single headed partial or full match, and as an extension
of an existing partial match into either a new (larger) partial match or a full
match. In order to extend partial matches, all previously computed matches are
stored. A scheduler decides which partial match is extended with which 
constraint, or which full match has its corresponding rule instance fired. More
details on the scheduler are given in Section~\ref{sub:new_meta:scheduler}.

First, to simplify the presentation, we propose an alternative syntax for 
\chrrp{} rules. An \emph{intermediate form} \chrrp{} rule looks as follows:
\[p :: r\ @\ s_1 A_1, \ldots, s_n A_n\iff B\]
where $s_i\in\{+,-,?\}$ and $A_i$ is an atom for $1\leq i\leq n$. If $s_i=+$
or $s_i=-$ then $A_i$ must be a CHR constraint and if $s_i=?$ then $A_i$ must
be a built-in constraint. An intermediate form \chrrp{} rule corresponds to
a regular \chrrp{} rule as follows: a term $+ A$ corresponds to a kept head
$A$, a term $- A$ corresponds to a removed head $A$, and a term $? A$ 
corresponds to a conjunct of the rule guard. The main advantage of the
intermediate form is that it supports specifying a join order for the heads, as
well as an evaluation order for the guards. In particular it supports 
specifying the evaluation of part of the guard after having computed only a
partial rule match. The intermediate form gives us the same syntactical 
flexibility as exists in the Logical Algorithms language where comparisons are 
interleaved with the (kept and removed) user-defined antecedents.

Consider, in general,
a simpagation rule of the form
\[p :: r\ @\ H_1,\ldots,H_i\backslash H_{i+1},\ldots,H_n\iff g\mid B\]
where the guard $g$ is a conjunction of atomic guards $g_1,\ldots,g_m$. We can
rewrite this rule in intermediate form syntax (amongst others) as follows:
\[p :: r\ @\ +H_1,\ldots, +H_i, -H_{i+1},\ldots,-H_n,?g_1,\ldots,?g_m
\iff B\]
In the following, we assume that all rules have the following form
\[p :: r\ @\ \pm H_1, ?g_1,\pm H_2,?g_2,\ldots,\pm H_n,?g_n\iff B\]
where $\pm$ means $+$ or $-$. Each $g_i$ ($1\leq i\leq n$) can be a
conjunction of primitive built-in constraints, and can in particular also be
equal to $\true$. The transformation from regular \chrrp{} syntax to 
intermediate form syntax can be done automatically using the above 
transformation schema, or by hand.

Using terminology similar to that of \cite{la}, we refer to a partial match, 
matching the heads $H_1,\ldots,H_i$ and satisfying the partial guard 
$g_1\wedge \ldots\wedge g_{i-1}$, as a \emph{suspended strong prefix firing}.
If also the partial guard $g_i$ is satisfied, we speak of a \emph{regular} (or 
\emph{non-suspended}) strong prefix firing. A constraint matching the next head
$H_{i+1}$ is called a \emph{prefix extension} of such a (regular) strong prefix
firing. A prefix firing that consists of all heads is (also) called a 
(suspended or regular) \emph{rule firing}. Here, a rule firing actually 
means a rule instance that is fireable. To avoid confusion, we refer to the 
actual firing of such a rule firing as \emph{firing a rule instance}.
Every prefix firing contains the left-most head and hence determines the rule
priority. In our implementation, we assume that all guards are monotone, i.e.,
once they are entailed by the built-in constraint store, they remain entailed
in any later
state. This is in fact required by the CHR operational (and declarative)
semantics, although most current CHR systems also support non-monotone (impure)
guards like for example \texttt{var}/1 in CHR on top of Prolog.
\subsection{Program-Dependent Part}\label{sec:new_implementation:pd}
The program-dependent part of our implementation (i.e., the part that depends
on the actual program to be implemented) consists of rules for
\begin{itemize}
\item generating a representation for CHR constraint occurrences and deleting
them when the represented constraint is removed;
\item generating and scheduling constraints representing prefix firings, 
prefix extensions and rule firings and deleting them when a constituent 
constraint is removed;
\item matching prefix firings with prefix extensions, firing rule instances,
and managing suspended prefix and rule firings.
\end{itemize} 
The different types of rules of the program-dependent part are illustrated by
using a running example program, namely Dijkstra's shortest path algorithm,
already given in the Logical Algorithms language in Example~\ref{ex:dijkstra}
and given here in \chrrp{} intermediate form syntax. To illustrate non-trivial
head matching, we have added a rule \texttt{d1} that removes simple loops
from the input graph.
\begin{Verbatim}[fontsize=\small,frame=single,commandchars=*\{\}]
    1 :: d1 @ -e(V,_,V), *:*, ?true            *:*,           *; <=> true.
    1 :: d2 @ +source(V),*:*, ?true            *:*,           *; <=> dist(V,0).
    1 :: d3 @ -dist(V,D*vrbsubt{1}), ?true, +dist(V,D*vrbsubt{2}), ?(D*vrbsubt{2} < D*vrbsubt{1}) <=> true.
D + 2 :: d4 @ +dist(V,D),*:*, ?true, +e(V,C,U),*:*,  ?true    *; <=> dist(U,D+C).
\end{Verbatim}
\subsubsection{Constraint Occurrence Representation}
\label{sec:new_implementation:pd:occurrence}
Although \chrrp{} constraints and CHR constraints obviously have the same
syntax and semantics (i.e., multi-set semantics with non-monotone deletion), we
introduce a new representation for them to allow unambiguous reference, reduce
work in case of constraint reactivation, and support the efficient deletion of
those prefix firings, prefix extensions, and rule firings in which they 
participate (see further). For each \chrrp{} constraint of predicate $c/n$, we
create a set of unique \emph{occurrence representations}
$c$\verb!_occ_!$i/(n+1)$, one for each occurrence of the predicate in a rule
head. The arguments of a $c$\verb!_occ_!$i/(n+1)$ constraint consist of 
the arguments of the original $c/n$ constraint, together with a unique 
constraint identifier that is shared by all occurrence representations. This
identifier is an uninstantiated variable as long as the constraint is
in the store and is instantiated the moment that the constraint is
to be deleted. For each user-defined constraint predicate $c/n$ with $m$ 
occurrences, the occurrence representations are generated using rules of
the following form.
\begin{Verbatim}[commandchars=*\{\},fontsize=\small]
*vrb{c}(X*vrbsubt{1},*ldots,X*vrbsub{}{n}) <=> *vrb{c}_occ_1(X*vrbsubt{1},*ldots,X*vrbsub{}{n},Id), *ldots, *vrb{c}_occ_*vrb{m}(X*vrbsubt{1},*ldots,X*vrbsub{}{n},Id).
\end{Verbatim}
For the example program, these rules look as follows.
\begin{Verbatim}[fontsize=\small,frame=single,commandchars=*\{\}]
source(V) <=> source_occ_1(V,Id).
dist(V,D) <=> dist_occ_1(V,D,Id), dist_occ_2(V,D,Id), dist_occ_3(V,D,Id).
e(V,C,U)  <=> e_occ_1(V,C,U,Id),  e_occ_2(V,C,U,Id).
\end{Verbatim}
\subsubsection{RETE Memory Constraints}\label{sec:new_implementation:pd:rete}
Regular and suspended prefix firings as well as prefix extensions are 
represented as CHR constraints. We call them RETE memory constraints because 
they coincide with the alpha and beta memories of the RETE algorithm. The RETE
memory constraints contain all arguments of their constituent CHR constraints, 
as well as their identifiers. Each RETE memory constraint moreover has its own 
unique identifier. We use the following functors for RETE memory constraints:
\begin{itemize}
\item $r$\verb!_pf_!$i$ for a regular (non-suspended) prefix firing of rule 
$r$, consisting
of $i$ heads, and $r$\verb!_pf_!$i$\verb!_suspended! for its suspended version 
\item $r$\verb!_pe_!$i$ for a prefix extension, consisting of the \ith{i+1} 
head of rule $r$
\item $r$\verb!_rf! for a (regular) rule firing of rule $r$ and
$r$\verb!_rf_suspended! for its suspended version.
\end{itemize}
If in a rule $r$, the partial guard after the \ith{i} head equals \true, then
there is no suspended version of the $i$-headed prefix firings of $r$, or of 
its rule firings if $r$ is an $i$-headed rule. In the example program, the 
following prefix firings, prefix extensions and rule firings are defined:
\begin{itemize}
\item \verb!d1_rf!/4
\item \verb!d2_rf!/3
\item \verb!d3_pf_1!/4, \verb!d3_pe_1!/3, \verb!d3_rf!/6 and 
    \verb!d3_rf_suspended!/6
\item \verb!d4_pf_1!/4, \verb!d4_pe_1!/4 and \verb!d4_rf!/7
\end{itemize}
\subsubsection{Suspended Prefix and Rule Firings}
\label{sec:new_implementation:pd:suspended}
Suspended prefix and rule firings are converted into regular prefix and rule
firings as soon as the relevant part of the guard is entailed. If on the other
hand this partial guard is disentailed, the suspended prefix or rule firing is
removed. Given a rule in intermediate form syntax
\[p :: r\ @\ \pm H_1, ?g_1,\pm H_2,?g_2,\ldots,\pm H_n,?g_n\iff B\]
we generate the following rules:
\begin{itemize}
\item For each $i$-headed suspended prefix firing:
\begin{Verbatim}[commandchars=*\{\},fontsize=\small]
*vrb{r}_pf_*vrb{i}_suspended(X*vrbsubt{1},*ldots,X*vrbsub{}{m},Id*vrbsubt{1},*ldots,Id*vrbsub{}{i},SId) <=> 
            *vrbsub{g}{i} | *vrb{r}_pf_*vrb{i}(X*vrbsubt{1},*ldots,X*vrbsub{}{m},Id*vrbsubt{1},*ldots,Id*vrbsub{}{i},SId),
            schedule_pf(*vrb{r}_*vrb{i}(Y*vrbsubt{1},*ldots,Y*vrbsub{}{l}),*vrb{p},SId).
*vrb{r}_pf_*vrb{i}_suspended(X*vrbsubt{1},*ldots,X*vrbsub{}{m},Id*vrbsubt{1},*ldots,Id*vrbsub{}{i},SId) <=> \+ *vrbsub{g}{i} | true.
\end{Verbatim}
where \verb!Y!\vrbsubt{1}$,\ldots,$\verb!Y!$_l$ are those variables in 
\verb!X!\vrbsubt{1}$,\ldots,$\verb!X!$_m$ that also appear in $H_{i+1}$
\item For each rule firing:
\begin{Verbatim}[commandchars=*\{\},fontsize=\small]
*vrb{r}_rf_suspended(X*vrbsubt{1},*ldots,X*vrbsub{}{m},Id*vrbsubt{1},*ldots,Id*vrbsub{}{n},SId) <=> *vrbsub{g}{n} |
            *vrb{r}_rf(X*vrbsubt{1},*ldots,X*vrbsub{}{m},Id*vrbsubt{1},*ldots,Id*vrbsub{}{n},SId), schedule_rf(*vrb{p},SId).
*vrb{r}_rf_suspended(X*vrbsubt{1},*ldots,X*vrbsub{}{m},Id*vrbsubt{1},*ldots,Id*vrbsub{}{n},SId) <=> \+ *vrbsub{g}{n} | true.
\end{Verbatim}
\end{itemize}
Note that if $g_i$ or $g_n$ equals $\true$, then we can apply unfolding to
replace occurrences of respectively 
$r$\verb!_pf_!$i$\verb!_suspended!/$(m+i+1)$
and $r$\verb!_rf_suspended!/$(m+n+1)$ by the bodies of the corresponding rules
above (see \cite{unfolding}). After this unfolding step, some of the above
rules may be removed. In the example program, only a rule firing of rule
\verb!d3! can be suspended. The code below is generated for such a rule firing.
\begin{Verbatim}[fontsize=\small,frame=single,commandchars=*\{\}]
d3_rf_suspended(V,D*vrbsubt{1},D*vrbsubt{2},Id*vrbsubt{1},Id*vrbsubt{2},SId) <=> 
	D*vrbsubt{2} < D*vrbsubt{1} | d3_rf(V,D*vrbsubt{1},D*vrbsubt{2},Id*vrbsubt{1},Id*vrbsubt{2},SId), schedule_rf(1,SId).
d3_rf_suspended(V,D*vrbsubt{1},D*vrbsubt{2},Id*vrbsubt{1},Id*vrbsubt{2},SId) <=> \+ (D*vrbsubt{2} < D*vrbsubt{1}) | true.
\end{Verbatim}
In the second rule above, \verb!\+ (!$C$\verb!)! is a safe approximation of the
negation
of constraint $C$, i.e., it is only entailed if constraint $C$ cannot possibly
hold. In the Prolog context, the built-in negation as failure can be used.

Suspended constraints are attached to all guarded variables so that they are 
reactivated whenever one of these variables is affected by a built-in 
constraint. We assume that both attaching and detaching
can be done in constant time, although certain current CHR implementations like
the K.U.Leuven CHR system do not support detaching in constant time.
\subsubsection{Scheduling}\label{sec:new_implementation:pd:scheduling}
Each constraint occurrence corresponds to a (potentially suspended) rule firing
if it is the only
head of a single headed rule, a (potentially suspended) prefix firing if it
is the first head of a multi-headed rule, and a prefix extension in all other
cases. A conversion between constraint occurrence and rule firing, prefix
firing or prefix extension is made as soon as the constraint in question 
matches with the head. If such a match is shown to be impossible, the 
constraint occurrence is discarded. 
Let there be given a head constraint $c(X_1,\ldots,X_n)$. The following 
function is used to construct a head match.
\begin{align*}
\headmatch([X|\bar{X}])=&
\begin{cases}
\langle [X|\bar{Y}],g\rangle&\textrm{if $X$ is a variable and 
    $X\notin vars(\bar{X})$}\\
\langle [Y|\bar{Y}],(Y=X)\wedge g\rangle&\textrm{otherwise}
\end{cases}\\
&\textrm{where $\langle \bar{Y},g\rangle = \headmatch(\bar{X})$}\\
\headmatch([]) =&\langle [],\true\rangle
\end{align*}
Now, for each rule in intermediate form syntax
\[p :: r\ @\ \pm H_1, ?g_1,\pm H_2,?g_2,\ldots,\pm H_n,?g_n\iff B\]
and for $1\leq i\leq n$ we generate the rules below
where $H_i=c(X'_1,\ldots,X'_n)$ is the \ith{j} occurrence of the 
user-defined constraint 
predicate $c/n$, $\langle [$\verb!X!$_1,\ldots,$\verb!X!$_n],g\rangle=
\headmatch([X'_1,\ldots,X'_n])$, and 
$\{$\verb!Y!$_1,\ldots,$\verb!Y!$_m\}=vars(H_i)\setminus
vars(\{H_1,\ldots,H_{i-1}\})$.
\begin{itemize}
\item If $i=n=1$:
\begin{Verbatim}[commandchars=*\{\},fontsize=\small]
*vrb{c}_occ_*vrb{j}(X*vrbsubt{1},*ldots,X*vrbsub{}{n},Id) <=> *vrb{g} | *vrb{r}_rf_suspended(Y*vrbsubt{1},*ldots,Y*vrbsub{}{m},Id,SId).
*vrb{c}_occ_*vrb{j}(X*vrbsubt{1},*ldots,X*vrbsub{}{n},Id) <=> \+ *vrb{g} | true.
\end{Verbatim}
\item If $i=1$ and $n>1$:
\begin{Verbatim}[commandchars=*\{\},fontsize=\small]
*vrb{c}_occ_*vrb{j}(X*vrbsubt{1},*ldots,X*vrbsub{}{n},Id) <=> *vrb{g} | *vrb{r}_pf_1_suspended(Y*vrbsubt{1},*ldots,Y*vrbsub{}{m},Id,SId).
*vrb{c}_occ_*vrb{j}(X*vrbsubt{1},*ldots,X*vrbsub{}{n},Id) <=> \+ *vrb{g} | true.
\end{Verbatim}
\item Otherwise,
 if $i>1$:
\begin{Verbatim}[commandchars=*\{\},fontsize=\small]
*vrb{c}_occ_*vrb{j}(X*vrbsubt{1},*ldots,X*vrbsub{}{n},Id) <=> *vrb{g} | 
      *vrb{r}_pe_*vrb{i-1}(Y*vrbsubt{1},*ldots,Y*vrbsub{}{m},Id,SId), schedule_pe(*vrb{r}_*vrb{i-1}(Z*vrbsubt{1},*ldots,Z*vrbsub{}{l}),SId).
*vrb{c}_occ_*vrb{j}(X*vrbsubt{1},*ldots,X*vrbsub{}{n},Id) <=> \+ *vrb{g} | true.
\end{Verbatim}
where $\{$\verb!Z!\vrbsubt{1}$,\ldots,$\verb!Z!$_l\}=vars(H_i)\cap
vars(\{H_1,\ldots,H_{i-1}\})$.
\end{itemize}
In the above, if $g=\true$ then the second rule of each pair of rules can be
discarded. The suspended prefix and rule firings can sometimes be replaced by
regular prefix and rule firings by unfolding (see 
Section~\ref{sec:new_implementation:pd:suspended}).

In the example program, only the first
occurrence of the $e/3$ constraint has a non-trivial head match (the first and
last argument must be the same). All prefix and rule firings are followed by 
the trivial guard $\true$ and so we only generate regular prefix and rule
firings. They are scheduled using the \verb!schedule_pf!/3 and 
\verb!schedule_rf!/2 predicates. 
\begin{Verbatim}[frame=single,fontsize=\small]
source_occ_1(V,Id) <=> d2_rf(V,Id,SId), schedule_rf(1,SId).

dist_occ_1(V,D,Id) <=> d3_pf_1(V,D,Id,SId), schedule_pf(d3_1(V),1,SId).
dist_occ_2(V,D,Id) <=> d3_pe_1(D,Id,SId), schedule_pe(d3_1(V),SId).
dist_occ_3(V,D,Id) <=> d4_pf_1(V,D,Id,SId), schedule_pf(d4_1(V),D+2,SId).

e_occ_1(V,C,U,Id) <=> V = U | d1_rf(V,C,Id,SId), schedule_rf(1,SId).
e_occ_1(V,C,U,Id) <=> \+ (V = U) | true.
e_occ_2(V,C,U,Id) <=> d4_pe_1(C,U,Id,SId), schedule_pe(d3_1(V),SId).
\end{Verbatim}
Prefix firings and extensions are scheduled using a key containing their shared
variables. For example for the prefix firings consisting of the
first head of rule \texttt{d3} and the corresponding prefix extensions 
consisting of the second head of the same rule, the key equals \verb!d3_1(V)!.

Similar to the suspended prefix and rule firings, the constraint occurrences
are attached to all guarded variables. We again assume that both attaching and
detaching can be done in constant time.
\subsubsection{Matching and Firing}
\label{sec:new_implementation:pd:match_and_fire}
The scheduler initiates the firing of a rule instance by asserting a 
\verb!fire!/1 constraint, and the matching of a prefix firing with a prefix
extension by asserting a \verb!match!/2 constraint. These constraints have as
arguments the identifiers of the corresponding RETE memory constraints. After
matching a prefix firing with a prefix extension, a new suspended prefix or
rule firing is generated. For a given $n$-headed rule $r$ with 
$n>1$ and for $1\leq i\leq n-2$, we generate the following rule
\begin{Verbatim}[commandchars=*\{\},fontsize=\small]
*vrb{r}_pf_*vrb{i}(X*vrbsubt{1},*ldots,X*vrbsub{}{m},Id*vrbsubt{1},*ldots,Id*vrbsub{}{i},SId*vrbsubt{1}), *vrb{r}_pe_*vrbit{i}(X*vrbsub{}{m+1},*ldots,X*vrbsub{}{l},Id*vrbsub{}{i+1},SId*vrbsubt{2}) \ 
        match(SId*vrbsubt{1},SId*vrbsubt{2}) <=> Id*vrbsub{}{i+1} \== Id*vrbsubt{1}, *ldots, Id*vrbsub{}{i+1} \== Id*vrbsub{}{i} |
                *vrb{r}_pf_*vrb{i+1}_suspended(X*vrbsubt{1},*ldots,X*vrbsub{}{l},Id*vrbsubt{1},*ldots,Id*vrbsub{}{i+1}).
\end{Verbatim}
and similarly for $i=n-1$:
\begin{Verbatim}[commandchars=*\{\},fontsize=\small]
*vrb{r}_pf_*vrb{n-1}(X*vrbsubt{1},*ldots,X*vrbsub{}{m},Id*vrbsubt{1},*ldots,Id*vrbsub{}{n-1},SId*vrbsubt{1}), *vrb{r}_pe_*vrbit{n-1}(X*vrbsub{}{m+1},*ldots,X*vrbsub{}{l},Id*vrbsub{}{n},SId*vrbsubt{2}) \ 
        match(SId*vrbsubt{1},SId*vrbsubt{2}) <=> Id*vrbsub{}{n} \== Id*vrbsubt{1}, *ldots, Id*vrbsub{}{n} \== Id*vrbsub{}{n-1} |
                *vrb{r}_rf_suspended(X*vrbsubt{1},*ldots,X*vrbsub{}{l},Id*vrbsubt{1},*ldots,Id*vrbsub{}{n}).
\end{Verbatim}
A rule firing of an $n$-headed rule $r$ with body $B$ is fired as follows:
\begin{Verbatim}[commandchars=*\{\},fontsize=\small]
*vrb{r}_rf_*vrb{i}(X*vrbsubt{1},*ldots,X*vrbsub{}{m},Id*vrbsubt{1},*ldots,Id*vrbsub{}{n},SId), fire(SId) <=> 
                Id*vrbsub{}{r(1)} = dead, *ldots, Id*vrbsub{}{r(l)} = dead, *vrb{B}.
\end{Verbatim}
where $r(1),\ldots,r(l)$ are the indices of the removed heads of the rule (if 
any). We furthermore add the following rules at the end of the code, to make 
sure the CHR compiler detects that the \texttt{match}/2 and \texttt{fire}/1
constraints are never to be stored.
\begin{verbatim}
match(_,_) <=> true.
fire(_) <=> true.
\end{verbatim}
For the example program, the generated code is as follows:
\begin{Verbatim}[fontsize=\small,frame=single,commandchars=*\{\}]
d1_rf(V,C,Id,SId), fire(SId) <=> Id = dead.
d2_rf(V,Id,SId), fire(SId) <=> dist(V,0).
d3_pf_1(V,D*vrbsubt{1},Id*vrbsubt{1},SId*vrbsubt{1}), d3_pe_1(D*vrbsubt{2},Id*vrbsubt{2},SId*vrbsubt{2}) \ match(SId*vrbsubt{1},SId*vrbsubt{2}) <=>
	Id*vrbsubt{2} \== Id*vrbsubt{1} | d3_rf_suspended(V,D*vrbsubt{1},D*vrbsubt{2},Id*vrbsubt{1},Id*vrbsubt{2},SId).
d3_rf(V,D*vrbsubt{1},D*vrbsubt{2},Id*vrbsubt{1},Id*vrbsubt{2},SId), fire(SId) <=> Id*vrbsubt{1} = dead.
d4_pf_1(V,D,Id*vrbsubt{1},SId*vrbsubt{1}), d4_pe_1(C,U,Id*vrbsubt{2},SId*vrbsubt{2}) \ match(SId*vrbsubt{1},SId*vrbsubt{2}) <=>
	Id*vrbsubt{2} \== Id*vrbsubt{1} | d4_pf(V,D,C,U,Id*vrbsubt{1},Id*vrbsubt{2},SId), schedule_rf(D+2,SId).
d4_rf(V,D,C,U,Id*vrbsubt{1},Id*vrbsubt{2},SId), fire(SId) <=> dist(U,D+C).

match(_,_) <=> true.
fire(_) <=> true.
\end{Verbatim}

\subsubsection{Clean-up}\label{sec:new_implementation:pd:clean}
Whenever a constraint's identifier variable is instantiated, its occurrence
representations, as well as those RETE memory constraints in which it 
participates, are removed. The rules look as follows.
\begin{itemize}
\item For the \ith{i} occurrence representation for constraint predicate $c/n$:
\begin{Verbatim}[commandchars=*\{\},fontsize=\small]
*vrb{c}_occ_*vrb{i}(X*vrbsubt{1},*ldots,X*vrbsub{}{n},Id) <=> nonvar(Id) | true.
\end{Verbatim}
\item For an $i$-headed suspended prefix firing of rule $r$:
\begin{Verbatim}[commandchars=*\{\},fontsize=\small]
*vrb{r}_pf_*vrb{i}_suspended(X*vrbsubt{1},*ldots,X*vrbsub{}{m},Id*vrbsubt{1},*ldots,Id*vrbsub{}{i},SId) <=> nonvar(Id*vrbsubt{1}) | true.
*ldots
*vrb{r}_pf_*vrb{i}_suspended(X*vrbsubt{1},*ldots,X*vrbsub{}{m},Id*vrbsubt{1},*ldots,Id*vrbsub{}{i},SId) <=> nonvar(Id*vrbsubt{i}) | true.
\end{Verbatim}
\item For an $i$-headed regular prefix firing of rule $r$:
\begin{Verbatim}[commandchars=*\{\},fontsize=\small]
*vrb{r}_pf_*vrb{i}(X*vrbsubt{1},*ldots,X*vrbsub{}{m},Id*vrbsubt{1},*ldots,Id*vrbsub{}{i},SId) <=> nonvar(Id*vrbsubt{1}) | remove_pf(SId).
*ldots
*vrb{r}_pf_*vrb{i}(X*vrbsubt{1},*ldots,X*vrbsub{}{m},Id*vrbsubt{1},*ldots,Id*vrbsub{}{i},SId) <=> nonvar(Id*vrbsubt{i}) | remove_pf(SId).
\end{Verbatim}
\item For a prefix extension of an $i$-headed prefix firing of rule $r$:
\begin{Verbatim}[commandchars=*\{\},fontsize=\small]
*vrb{r}_pe_*vrb{i}(X*vrbsubt{1},*ldots,X*vrbsub{}{m},Id,SId) <=> nonvar(Id) | remove_pe(SId).
\end{Verbatim}
\item For a suspended rule firing of an $n$-headed rule $r$:
\begin{Verbatim}[commandchars=*\{\},fontsize=\small]
*vrb{r}_rf_suspended(X*vrbsubt{1},*ldots,X*vrbsub{}{m},Id*vrbsubt{1},*ldots,Id*vrbsub{}{n},SId) <=> nonvar(Id*vrbsubt{1}) | true.
*ldots
*vrb{r}_rf_suspended(X*vrbsubt{1},*ldots,X*vrbsub{}{m},Id*vrbsubt{1},*ldots,Id*vrbsub{}{n},SId) <=> nonvar(Id*vrbsubt{n}) | true.
\end{Verbatim}
\item For a regular rule firing of an $n$-headed rule $r$: 
\begin{Verbatim}[commandchars=*\{\},fontsize=\small]
*vrb{r}_rf(X*vrbsubt{1},*ldots,X*vrbsub{}{m},Id*vrbsubt{1},*ldots,Id*vrbsub{}{n},SId) <=> nonvar(Id*vrbsubt{1}) | remove_rf(SId).
*ldots
*vrb{r}_rf(X*vrbsubt{1},*ldots,X*vrbsub{}{m},Id*vrbsubt{1},*ldots,Id*vrbsub{}{n},SId) <=> nonvar(Id*vrbsubt{n}) | remove_rf(SId).
\end{Verbatim}
\end{itemize}
The predicates \verb!remove_pf!/1, \verb!remove_pe!/1 and \verb!remove_rf!/1
remove respectively a prefix firing, prefix extension and rule firing from
the schedule. The following clean-up rules are generated for the example
program.
\begin{Verbatim}[fontsize=\small,frame=single,commandchars=*\{\}]
source_occ_1(V,Id) <=> nonvar(Id) | true.
dist_occ_1(V,D,Id) <=> nonvar(Id) | true.
dist_occ_2(V,D,Id) <=> nonvar(Id) | true.
dist_occ_3(V,D,Id) <=> nonvar(Id) | true.
e_occ_1(V,C,U,Id)  <=> nonvar(Id) | true.
e_occ_2(V,C,U,Id)  <=> nonvar(Id) | true.

d1_rf(V,C,Id,SId)        *vrbsubt{* }*vrbsubt{* }<=> nonvar(Id)*vrbsubt{* } | remove_rf(SId).
d2_rf(V,Id,SId)          *vrbsubt{* }*vrbsubt{* }<=> nonvar(Id)*vrbsubt{* } | remove_rf(SId).
d3_pf_1(V,D*vrbsubt{1},Id*vrbsubt{1},SId)      <=> nonvar(Id*vrbsubt{1}) | remove_pf(SId).
d3_pe_1(D*vrbsubt{2},Id*vrbsubt{2},SId)        <=> nonvar(Id*vrbsubt{2}) | remove_pe(SId).
d3_rf(V,D*vrbsubt{1},D*vrbsubt{2},Id*vrbsubt{1},Id*vrbsubt{2},SId) *:<=> nonvar(Id*vrbsubt{1}) | remove_rf(SId).
d3_rf(V,D*vrbsubt{1},D*vrbsubt{2},Id*vrbsubt{1},Id*vrbsubt{2},SId) *:<=> nonvar(Id*vrbsubt{2}) | remove_rf(SId).
d4_pf_1(V,D,Id*vrbsubt{1},SId)    *vrbsubt{* }  <=> nonvar(Id*vrbsubt{1}) | remove_pf(SId).
d4_pe_1(C,U,Id*vrbsubt{2},SId)    *vrbsubt{* }  <=> nonvar(Id*vrbsubt{2}) | remove_pe(SId).
d4_rf(V,D,C,U,Id*vrbsubt{1},Id*vrbsubt{2},SId) <=> nonvar(Id*vrbsubt{1}) | remove_rf(SId).
d4_rf(V,D,C,U,Id*vrbsubt{1},Id*vrbsubt{2},SId) <=> nonvar(Id*vrbsubt{2}) | remove_rf(SId).

d3_rf_suspended(V,D*vrbsubt{1},D*vrbsubt{2},Id*vrbsubt{1},Id*vrbsubt{2},SId) <=> nonvar(Id*vrbsubt{1}) | true.
d3_rf_suspended(V,D*vrbsubt{1},D*vrbsubt{2},Id*vrbsubt{1},Id*vrbsubt{2},SId) <=> nonvar(Id*vrbsubt{2}) | true.
\end{Verbatim}

\subsection{Program-Independent Part: the Scheduler}
\label{sub:new_meta:scheduler}
The scheduler implements the \verb!schedule_rf!/2, \verb!remove_rf!/1,
\verb!schedule_pf!/3,\linebreak \verb!remove_pf!/1, \verb!schedule_pe!/2
and \verb!remove_pe!/1 predicates.
It furthermore implements the \verb!execute!/0
predicate which retrieves and executes the highest priority scheduled task.
This task either is the firing of a rule instance by asserting a \verb!fire!/1 
constraint, or the matching of a prefix firing with a prefix extension by
asserting a \verb!match!/2 constraint. The \verb!execute!/0 predicate 
recursively calls itself until no more tasks are scheduled. It is first
called after processing the initial goal.

For the implementation of the scheduler, we use a variant of the scheduling 
algorithm presented in \cite{mergeable_schedules}. This algorithm can be used
to maintain which prefix firings are still to match with which prefix
extensions. It is roughly based on the $\mathcal{W}(r,t)$ data structures used
in \cite{la}. Such a data structure consists of a series (implemented as a 
linear
linked list) of \emph{prefix blocks}, which are sets of prefix firings and
(apart from the last one) are associated with a prefix extension. 

The semantics of the $\mathcal{W}(r,t)$ data structure is that the
prefix firings of a given prefix block are still to match with the prefix
extension associated to it, as well as with all prefix extensions associated to 
subsequent prefix blocks. The last prefix block has no associated prefix 
extension, and represents those prefix firings that have been matched with
all prefix extensions and hence are passive (or \emph{completed} using the 
terminology of \cite{la}). Whenever a prefix extension is deleted, its prefix
block is merged with the next prefix block.

There is one $\mathcal{W}(r,t)$ data structure for each prefix length of
each rule and for each combination of arguments shared between a prefix
firing and prefix extension. Each prefix block is represented as a (local)
priority queue whose items are the block's prefix firing. The highest
priority item of each prefix block, together with its associated prefix 
extension, is also represented in a global priority queue. This prefix block
representative is updated whenever the highest priority prefix firing of the
prefix block is removed, a new prefix firing has the highest priority, or the
associated prefix extension is removed. The global priority queue furthermore 
contains a representative for each rule firing. The reason for using
two layers of priority queues is to reduce the amount of work needed when the 
prefix firings of a prefix block all become passive due to a prefix extension
removal. It is the global priority queue that determines the next task to
perform, i.e., matching a prefix firing with a prefix extension, or firing a
rule instance.

In the context of \chrrp{}, built-in constraint (in particular equality
constraints) on the arguments shared between a prefix firing and extension,
may require merging of $\mathcal{W}(r,t)$ data structures. The data structure
of \cite{mergeable_schedules} supports schedule merges in quasi constant time.
The most notable difference with the $\mathcal{W}(r,t)$ data structure of 
\cite{la} is that the prefix blocks form a circular linked list. Using this
representation, merging schedules consists of cross-linking the circular lists
and reactivating the prefix firings that were passive before the merge.
Special care is taken to prevent both that a prefix firing is being 
matched with the same prefix extension more than once, and that a prefix 
firing `misses' a prefix extension. 

One consequence of using a circular linked list instead of a linear one to 
represent the prefix blocks, is that it is unclear (or more precisely, too 
expensive to decide) which prefix firings become passive whenever a prefix
extension is deleted. Therefore, this decision is postponed until the scheduler
tries to match the prefix firing with
the next prefix extension in line. For complexity reasons, it is important that
all prefix firings that have simultaneously been reactivated, and have not
been matched with a prefix extension since this reactivation, are 
simultaneously made passive in time independent of the number of prefix firings
affected. In \cite{mergeable_schedules}, a so-called \emph{element schedule}
based on a stack is proposed to supports this. In our context, we need an
element schedule that is based on priority queues. It works as follows.

We use three types of priority queues. The first one is a single \emph{global}
priority queue which contains an item for each rule firing, for each 
\emph{active} prefix firings that either has not been passive before or has 
been matched with at least one prefix extension since its last activation, and
finally, for each set of prefix firings that have been simultaneously activated
and have not been matches with a prefix extension since. A second type of
priority queues is called a \emph{local} queue and represents the above 
mentioned sets of prefix firings. Finally, the third type of queues is the 
\emph{passive} queue which contains an item for each passive (completed) prefix
firing. There is one passive queue for each schedule. Essentially, we again use
two layers of priority queues. Whenever a set of previously passive prefix 
firings, represented as a passive priority queue, is reactivated because
of a new prefix extension or because of a schedule merge, this passive priority
queue becomes a local priority queue and has a representative 
inserted into the global priority queue. If such a representative is the 
highest priority item in the global priority queue, and an \verb!execute!/0 
call is made, then the highest priority prefix firing of the represented local
priority is removed and dealt with as an ordinary prefix firing. The 
representatives of local priority queues are updated (and potentially removed)
similarly to how this is done in the $\mathcal{W}(r,t)$ data structure of
\cite{la}.
\begin{example}
Figure~\ref{fig:queues} illustrates the prefix blocks, the different types of 
priority queues, and their contents.
\begin{figure}
\begin{center}  
\includegraphics[width=0.6067\linewidth,keepaspectratio]{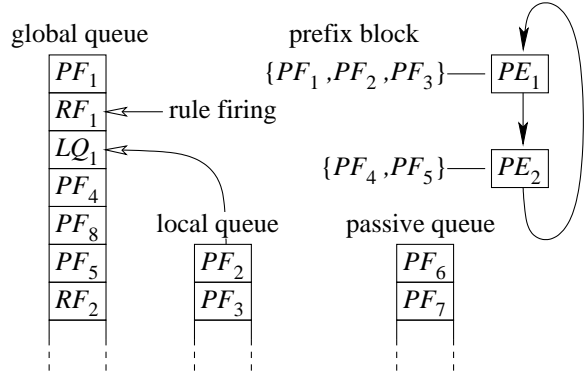}
\end{center}
\caption{Example schedule with global, local and passive priority queues
\label{fig:queues}}
\end{figure}
The global queue, which is shared by all schedules, contains the rule firings
$\mathit{RF}_1$ and $\mathit{RF}_2$, the prefix firings $\mathit{PF}_1$,
$\mathit{PF}_4$, $\mathit{PF}_5$ and $\mathit{PF}_8$ (the last of which belongs
to another schedule), and the local queue representative $\mathit{LQ}_1$. The
represented
local queue contains the prefix firings $\mathit{PF}_2$ and $\mathit{PF}_3$ 
which are by definition also in the same prefix block. The schedule's passive
queue contains the prefix firings $\mathit{PF}_6$ and $\mathit{PF}_7$. The 
schedule has two prefix blocks, which are associated with respectively the
prefix extensions $\mathit{PE}_1$ and $\mathit{PE}_2$.
\end{example}
Using our approach, the cost of deleting items from the global priority queue 
can be amortized to one of the following events: a new rule firing, a new 
prefix firing, a new prefix extension (for each representative of a local 
priority queue), or a match between a prefix firing and a prefix extension 
(which corresponds to either a new larger prefix firing, or a rule firing).  

In \cite{la}, retrieving the schedule for a given prefix firing or prefix
extension is done by hashing. In our approach, we use a variant of hashing,
which we call \emph{non-ground hashing} and which consists of first replacing
all variables by a unique identifier, and then using the resulting (ground)
term for hashing. Unifications may require rehashing the affected keys and 
potentially also the merging of schedules.
\subsection{Priority Queues}\label{sec:new_implementation:priority}
A priority queue or heap is a data structure that contains a set of prioritized
items and supports the following operations: inserting and removing an item,
finding a highest priority item and merging with another queue. The
implementation proposal in \cite{la} suggests the use of two types of priority
queues, one for the fixed priorities, where each of the supported operations
takes constant time, and Fibonacci heaps for the dynamic priorities.

Fibonacci heaps \cite{tarjan} are a type of priority queue that offer
$\mathcal{O}(1)$ amortized time insertion, heap merging and finding a highest
priority item, and $\mathcal{O}(\log n)$ amortized time item removal with $n$
the number of items in the queue. It is suggested in \cite{la} that by using
only one node per priority, using linked lists to represent the items that
share this priority, the item removal cost can be reduced to $\mathcal{O}(\log
N)$ with $N$ the number of distinct priorities. However, this increases the
cost of heap
merging from $\mathcal{O}(1)$ for a single merge operation to a total cost of 
$\mathcal{O}(n\log N)$ for merging heaps when there are $n$ items in total and
$N$ distinct priorities (as is shown in an Appendix of \cite{chrla}).
A CHR implementation of Fibonacci heaps is described
in \cite{jon:dijkstra}. It can easily be extended to support multiple heaps
that can be merged and to use only one node for each distinct priority per
heap.
\section{A New Meta-Complexity Result for \chrrp}\label{sec:new_meta}
In this section, we give a new meta-complexity result for \chrrp{}. It extends
the result via translation to Logical Algorithms, by also supporting built-in
constraints and non-ground CHR constraints. We make the following assumptions:
\begin{itemize}
\item Hash tables support $\mathcal{O}(1)$ insertion, removal, and retrieval of all
elements that match a given (ground) key.
\item The inverse of the Ackermann function ($\alpha(n)$) is a constant.
\end{itemize}
The first assumption is also made in \cite{la} and holds on average as long
as the hash function is good enough. The second assumption is needed for our
scheduling data structure \cite{mergeable_schedules} which internally makes use
of the union-find algorithm. The inverse of the Ackermann function is positive
and less than 5 for all practical purposes.

We start by looking at the complexity of the different operations supported by
our scheduler.
\begin{lemma}[Scheduler Costs]\label{lem:schedule_costs}
Let $N$ be the number of distinct priorities, and assume that a priority queue
merge takes some abstract time $T$, then the schedule operations have the 
following amortized cost:
\begin{itemize}
\item $\mathcal{O}(1)$ and $\mathcal{O}(\log N)$ for each \verb!schedule_pf!/3, 
\verb!remove_pf!/1, \verb!remove_pe!/1, \verb!remove_rf!/1 and
\verb!execute!/0 operation involving respectively a static and dynamic
priority rule
\item $\mathcal{O}(T+1)$ and $\mathcal{O}(T+\log N)$ for each 
\verb!schedule_pe!/2 operation involving respectively a static and dynamic
priority rule
\item $\mathcal{O}(1)$ for each schedule merge and \verb!schedule_rf!/2 
operation
\end{itemize}
\end{lemma}
\begin{proof}
We only consider the costs related to the priority queue operations. The other
costs are shown to be (quasi) constant in
\cite{mergeable_schedules}. We now look at the different operations in detail:
\begin{itemize}
\item A \verb!schedule_pf!/3 call consists of inserting the new prefix firing
into the global priority queue. We also account to this event, the cost of
making the new prefix instance passive the first time. That operation 
consists of a removal from the global priority queue and an insertion into the
schedule's passive queue. The total cost is $\mathcal{O}(1)$ if the element has
a static priority, and $\mathcal{O}(\log N)$ if it has a dynamic priority.
\item A \verb!schedule_pe!/2 call requires the insertion of a new 
representative for the local priority queue of reactivated prefix firings,
into the global priority queue. We also take into account here, the cost of
making all the reactivated prefix firings passive 
that have not been matched with a prefix
extension since the reactivation. That operation consists of removing the
representative and merging the local priority queue with the schedule's passive
queue. The cost is $\mathcal{O}(T+1)$ for a static priority rule and 
$\mathcal{O}(T+\log N)$ time for a dynamic priority one. 
\item A \verb!schedule_rf!/2 call requires an insertion into the global
priority queue which takes $\mathcal{O}(1)$ time.
\item A \verb!remove_pf!/1 call consists of deleting the prefix firing
from the global priority queue, from a local priority queue or from a passive
queue. A deletion from a local queue may moreover require an update of the
global queue (removal and insertion). In total, this takes $\mathcal{O}(1)$ 
time for a static priority rule and $\mathcal{O}(\log N)$ time for a dynamic
priority rule.
\item A \verb!remove_pe!/1 call does not require any priority queue operations,
and so the cost is $\mathcal{O}(1)$. 
\item A \verb!remove_rf!/1 call requires a removal from the global priority
queue which takes $\mathcal{O}(1)$ time if it involves a static priority rule
and $\mathcal{O}(\log N)$ time if it involves a dynamic priority rule.
\item An \verb!execute!/0 call requires retrieval and potential removal 
(if the retrieved item corresponds to a rule firing, or to a prefix firing that
becomes passive) of the highest priority item in the global priority queue. If
the retrieved item represents a prefix firing or set of prefix firings that
need to be made passive, the cost of this operation is already accounted for by
a previous \verb!schedule_pf!/3 or \verb!schedule_pe!/2 
operation. In such case, we call the \verb!execute!/0 call \emph{unsuccessful}.
An unsuccessful \verb!execute!/0 call is followed by another \verb!execute!/0
call until either such a call is successful, or the global priority queue is
empty and thus a final state is reached. The cost of all unsuccessful 
\verb!execute!/0 calls can be amortized to previous events. If in case of a
successful \verb!execute!/0 call, the item retrieved from the global priority
queue corresponds to the representative of a local priority queue, the 
operation requires a removal of the highest priority item (prefix firing) from
this local queue, an insertion of the prefix firing into the global priority
queue, and potentially the insertion of a new representative for the local
queue into the global queue. The cost of a successful \verb!execute!/0 call 
therefore equals $\mathcal{O}(1)$ if it involves a static priority rule and
$\mathcal{O}(\log N)$ otherwise.
\item A schedule merge requires the reactivation of the passive prefix firings
of the merged schedules. The cost analysis is similar to that of a
\verb!schedule_pe!/2 call. Moreover, each schedule merge can be accounted for
by at least one \verb!schedule_pe!/2 call as the resulting schedule contains
at least one prefix extension more than each of the original schedules, and so
the number of schedule merges is bounded by the number of prefix extensions.
Therefore, the cost of a single schedule merge can be considered constant.  
\end{itemize}
\end{proof}
In the above lemma, we have made abstraction of the cost of priority queue
merge operations. Such merges take place when the prefix firings in a local
priority queue all become passive. In such an event, the local priority queue
is merged with the schedule's passive queue. It is easy to see that the cost of
merging priority queues for static priorities takes constant time per merge 
operation. In Section~\ref{sec:new_implementation:priority}
a bound is given on the total cost of merging
Fibonacci heaps with one node per distinct priority, given the number of items
ever inserted into the heaps. The following lemma makes use of this result.
\begin{lemma}[Fibonacci Heap Merging Cost]
The total cost of Fibonacci heap merges is $\mathcal{O}((P_d+A_d)\cdot \log N)$
where $P_d$ is the number of strong prefix firings of dynamic priority rules,
$A_d$ is the number of constraints that may participate in a dynamic priority
rule instance, and $N$ is the number of distinct rule priorities.
\end{lemma}
\begin{proof}
We count the number of items ever inserted into the local and passive Fibonacci
heaps, and then apply the result of 
Section~\ref{sec:new_implementation:priority}. A local priority queue 
basically is the same as a passive priority queue in which items are no longer
inserted. Therefore, a merge between a local queue and a passive queue can be
seen as a special case of a merge between two passive queues and so we only
need to consider these passive priority queues. Each item inserted in such a
queue is either a prefix firing that has never been passive before, or a prefix
firing that has been matched with a prefix extension at least once since its
last activation. The total number of these items is $\mathcal{O}(P_d+A_d)$
because each prefix firing that has been matched with a prefix extension is by
definition a strong prefix firing, and each new prefix firing either
results from matching a (smaller) strong prefix firing and extension and hence
corresponds to a (potentially suspended) strong prefix firing, or 
consists of a single head in which case it corresponds to a constraint
assertion. Now given the number of items ever inserted into the passive 
priority queues, the total cost of merging Fibonacci heaps hence is
$\mathcal{O}((P_d+A_d)\cdot \log N))$. 
\end{proof}
We are now ready to formulate the new meta-complexity theorem.
\begin{theorem}\label{the:meta}
Let $A_s$ and $A_d$ be the number of assertions of constraints with an 
occurrence in respectively a static and dynamic priority rule. Let $P_s$ and
$P_d$ be the number of strong prefix firings of respectively static and
dynamic priority rules.
The time complexity of a \chrrp{} program executed using our implementation is
\[\mathcal{O}((1+C_B^\mathit{ask})\cdot(A_s+P_s+(A_d+P_d)\cdot\log N)+
B\cdot C_B^\mathit{tell}\cdot (K + C_B^\mathit{ask}\cdot S))\] where
$N$ is the number of distinct priorities, $C_B^\mathit{ask}$ is the cost of
evaluating a built-in ask constraint, $C_B^\mathit{tell}$ is the cost of 
solving a built-in
tell constraint, and $B$ is the number of built-in tell constraints asserted in
rule bodies; $K$ is the maximum number of distinct combinations (keys) of 
arguments shared between prefix firings and extensions in which any given
variable occurs, and $S$ is the maximum number of 
suspended strong prefix firings (i.e., those that are followed by a 
non-trivial guard) and suspended instances of constraint occurrences
(i.e., whose arguments are not mutually distinct variables) in which any given
variable occurs.
\end{theorem}
\begin{proof}
Each new CHR constraint causes the creation of constraint occurrences which
are converted into RETE memory constraints as soon as the implicit guard on
the constraint arguments is entailed (i.e., the constraint matches the head in
question). These RETE memory constraints are scheduled using 
\verb!schedule_pf!/3 for the single-headed prefix firings, \verb!schedule_rf!/2
for the single-headed rule firings, and \verb!schedule_pe!/2 for the prefix
extensions. The total cost of these operations, including the cost of priority
queue merges (for the \verb!schedule_pe!/2 calls), equals 
$\mathcal{O}((1+C_b^\mathit{ask})\cdot(A_s+(A_d+P_d)\log N))$. Each 
constraint deletion causes the deletion of those RETE memory constraints
in which the deleted constraint participated. The total cost related
to deletion therefore is $\mathcal{O}(A_s + P_s + (A_d + P_d)\log N)$.
Each prefix firing is inserted into its schedule at most once and hence it 
can also be removed from this schedule only once (when one of its constituent 
constraints is removed). Those prefix firings that consist of at least two
heads, correspond to a strong prefix firing as they are generated at a priority
higher or equal to that of the highest priority rule firing.
Thus, using Lemma~\ref{lem:schedule_costs} and including the cost of checking
the relevant parts of the guard, the cost for inserting (and deleting) these
prefix firings is $\mathcal{O}((1+C_b^\mathit{ask})\cdot(P_s+P_d\log N))$. 

A built-in tell constraint is processed as follows. The keys used to identify
the schedules and that are affected by the built-in constraint, are rehashed.
If the built-in constraint causes two or more schedules to have the same key, 
these schedules are merged. The cost of rehashing is proportional to the number
of affected keys and the cost of a schedule merge is constant by
Lemma~\ref{lem:schedule_costs}. A built-in constraint moreover requires the
reactivation of the suspended prefix firings and rule firings, as well as those
constraint occurrences for which it is not decided whether they match with the
corresponding head or not. The reactivated prefix and rule firings have their
guard checked and are potentially scheduled as regular (non-suspended) prefix
and rule firings. The reactivated constraint occurrences also have their 
(implicit) guard checked, and are potentially scheduled as single-headed prefix
firings, single-headed rule firings, or prefix extensions. The cost of the
scheduling operations was already taken into account above. The remaining cost
per built-in tell constraint is $\mathcal{O}(C_b^\mathit{tell}\cdot
(K+C_b^\mathit{ask}\cdot S))$. 
\end{proof}
The values of $S$ and $K$ might be difficult to determine in practice, but we
can use an upper bound of $\mathcal{O}(A_s+A_d+P_s+P_d)$ for both parameters.
The
reasoning for $S$ is that the number of suspended prefix firings is smaller
than the number of prefix firings and the number of suspended constraint
occurrences is smaller than the number of assertions times the number of rule
heads in the program. For $K$, we have that the number of keys shared between
prefix firings and extensions is limited by the total number of prefix firings
and extensions. We have used the cost of solving a 
built-in tell constraint as an upper bound on the number of variables that are
affected. 

The meta-complexity theorem also applies to (regular) CHR programs, which can
be seen as a special case of \chrrp{} programs in which all rules have the same
(static) priority; see Theorem 3 of \cite{chrrp} for more details. 
\subsection{Examples}\label{sub:new_meta:examples}
We illustrate the meta-complexity theorem on some examples, and compare with
the results obtained by using the approach of \cite{atgb2}.
\begin{example}[Less-or-Equal]\label{ex:new_meta:leq}
The less-or-equal (\texttt{leq}) program is classic CHR example. It implements
a less-than-or-equal-to constraint by eventually translating it into equality
constraints. A \chrrp{} implementation of the program consists of the following
rules.
\begin{verbatim}
1 :: idempotence  @ leq(X,Y) \ leq(X,Y) <=> true.
2 :: reflexivity  @ leq(X,X) <=> true.
2 :: antisymmetry @ leq(X,Y), leq(Y,X) <=> X = Y.
3 :: transitivity @ leq(X,Y), leq(Y,Z) ==> leq(X,Z).
\end{verbatim}
Given an initial goal consisting of $n$ \verb!leq!/2 constraints where the
arguments are taken from a set of $n$ distinct variables, we derive the 
following values for the parameters:
\begin{itemize}
\item $P_s$: the number of strong prefix firings is $\mathcal{O}(n^2)$ for the
\texttt{idempotence} rule, $\mathcal{O}(n)$ for the \texttt{reflexivity} rule,
$\mathcal{O}(n^2)$ for the \texttt{antisymmetry} rule, and
$\mathcal{O}(n^3)$ for the \texttt{transitivity} rule. These numbers are found
by looking at the degrees of freedom for each constraint occurrence, based on
the domain of the arguments, and given those arguments that are already fixed
by the left-most heads. For example for the \texttt{transitivity} rule, we
know that there are $\mathcal{O}(n^2)$ constraints matching the first head,
and $\mathcal{O}(n)$ constraints matching the second head, given the first. 
Our reasoning is based on the fact that at priority 2 and lower (numerically
larger), all
\texttt{leq}/2 constraints have set semantics because of the \verb!idempotence!
rule.
\item $A_s$: the number of \verb!leq!/2 constraints asserted is 
$\mathcal{O}(n^3)$ (by the \verb!transitivity! rule).
\item $B$: the number of built-in constraints is bounded by the number of rule
firings of the \texttt{antisymmetry} rule, and hence is $\mathcal{O}(n^2)$.
\item $K$: the schedule keys are the combination of \verb!X! and \verb!Y! in
both the \texttt{antisymmetry} rule and the \texttt{idempotence} rule, and 
\verb!Y! in the
\texttt{transitivity} rule. There are at most $\mathcal{O}(n)$ different keys
in which any given variable occurs.
\item $S$: for any variable, and in a state in which a built-in constraint can
be asserted, there are up to $\mathcal{O}(n)$ suspended instances of the
\verb!leq!/2 occurrence in the \verb!reflexivity! rule. There can be no
suspended prefix or rule firings.
\item $C_b^\mathit{ask}$ and $C_b^\mathit{tell}$: the cost of evaluating a
built-in ask 
constraint and the cost of solving a built-in tell constraint is constant (at
least for the given query pattern).
\end{itemize}
Filling in these parameters in the formula given by Theorem~\ref{the:meta} 
gives us a worst case time complexity of 
\[\mathcal{O}((1+1)\cdot(n^3+n^3+(0+0)\cdot\log 3)+n^2\cdot 1\cdot 
(n+1\cdot n))=\mathcal{O}(n^3)\]
This corresponds to the actual worst-case complexity for
an initial goal of the form 
\[\{\texttt{leq($X_1$,$X_2$)},\ldots,\texttt{leq($X_{n-1}$,$X_n$)},
\texttt{leq($X_n$,$X_1$)}\}\]
The approach of \cite{atgb2} does not apply since the
\texttt{transitivity} rule is a propagation rule and hence no suitable
ranking function can be found.\qed
\end{example} 
\begin{example}[Merge Sort]
Consider the \chrrp{} implementation of the merge sort algorithm, first given
in Example~\ref{ex:chrrp_to_la:merge_sort} (Section~\ref{chr-to-la}) and
repeated here for easy reference.
\begin{Verbatim}[frame=single,fontsize=\small]
1 :: ms1 @ arrow(X,A) \ arrow(X,B) <=> A < B | arrow(A,B).
2 :: ms2 @ merge(N,A),  merge(N,B) <=> A < B | merge(2*N+1,A), arrow(A,B).
3 :: ms3 @ number(X) <=> merge(0,X).
\end{Verbatim}
We show that the total runtime of the algorithm is $\mathcal{O}(n\log n)$ given
an initial goal consisting of $n$ \texttt{number}/1 constraints.

No new \texttt{number}/1 constraints are ever asserted. Rule 
\texttt{ms3} converts one \texttt{number}/1 constraint into one 
\texttt{merge}/2 constraint each time it fires. The number of (strong) prefix 
firings for rule \texttt{ms3} hence is $\mathcal{O}(n)$. 
Rule \texttt{ms2} decreases the number of \texttt{merge}/2 constraints by one
and so it can fire $n-1$ times. 
In any state, there are at most two \texttt{merge}/2 constraints with
the same first argument. This invariant holds in the initial state because
there are no \texttt{merge}/2 constraints in the initial goal and rule 
\texttt{ms2} can fire after each new \texttt{merge}/2 constraint assertion, 
enforcing the invariant. Because of the invariant, the number of prefix firings
for rule \texttt{ms2} is limited to $\mathcal{O}(n)$.

Using similar reasoning it holds that in any state, there are at most two
\texttt{arrow}/2 constraints in the store with the same first argument. Now
we define that in a given state, two numbers $X_1$ and $X_m$ are connected by a
\emph{chain} of length $m-1$ if the following constraints are in the store: 
\texttt{arrow($X_1$,$X_2$)}, \texttt{arrow($X_2$,$X_3$)}, \ldots,
\texttt{arrow($X_{m-1}$,$X_m$)}. 
At priority 2 it holds that for each \texttt{merge($N$,$X$)} constraint in the
store, the maximal length of a chain starting in $X$ is $N$. Indeed, this holds
for the initial \texttt{merge(0,\avar)} constraints and if it holds for 
\texttt{merge($N$,\avar)} constraints, it also holds for 
\texttt{merge($2\cdot N +1$,\avar)} constraints, because when such a constraint
is asserted, two chains of length $N$ are linked with an extra \texttt{arrow}/2
constraint and merged by up to $2\cdot N$ firings of rule \texttt{ms1}. 
Two \texttt{merge($N$,\avar)} constraints are combined into a 
\texttt{merge($2\cdot N+1$,\avar)} constraint, so the $n$ 
\texttt{merge(0,\avar)} constraints asserted by rule \texttt{ms3} are replaced
by $n/2$ \texttt{merge(1,\avar)} constraints, which in turn are combined into 
$n/4$ \texttt{merge(3,\avar)} constraints and so on until finally $1$ 
\texttt{merge($n-1$,\avar)} constraint remains. The sum of all $N$ in these 
\texttt{merge($N$,\avar)} constraints is $\mathcal{O}(n\log n)$.
Rule \texttt{ms1} fires $\mathcal{O}(N)$ times after every new 
\texttt{merge($N$,\avar)} constraint assertion and because there are at most 
two \texttt{arrow}/3 constraints with the same first argument, there are
$\mathcal{O}(n\log n)$ strong prefix firings of rule \texttt{ms1}.

In conclusion, for an initial goal consisting of $n$ 
\texttt{number}/1 constraints, there are $\mathcal{O}(n\log n)$ strong prefix
firings for rule \texttt{ms1}, $\mathcal{O}(n)$ for rule \texttt{ms2} and 
$\mathcal{O}(n)$ for rule \texttt{ms3}. Using the meta-complexity theorem, 
which simplifies to the one for Logical Algorithms because there are no
built-in tell constraints, the total runtime is $\mathcal{O}(n\log n)$, which
is also a tight complexity bound. We now compare this result with the result 
found by using the meta-complexity theorem of \cite{atgb2}.

Using a similar analysis as above, we can derive that $D=\mathcal{O}(n\log n)$
and $c_{max}=\mathcal{O}(n)$ where $n$ is the number of \texttt{number}/1 
constraints in the query. Note that in Theorem 4.2 of \cite{atgb2}, a worst
case upper bound of $c_{max}=\mathcal{O}(c+D)$ is used,
with $c$ the number of constraints in the query, which becomes 
$c_{max}=\mathcal{O}(n\log n)$ in this example. The bound we use is tight, 
i.e., $c_{max}=\Theta(n)$. 
The cost of head matching ($O_{H_r}$), guard checking ($O_{G_r}$), adding
built-in constraints ($O_{C_r}$), and adding and removing CHR constraints 
($O_{B_r})$, can all be assumed constant. The number of heads $n_r$ of a rule
$r\in P$ is at most 2.
Filling in these numbers, we derive a total worst case complexity of
$\mathcal{O}(n^3\log n)$, which is clearly suboptimal.
\end{example}
\begin{example}[Dijkstra's Shortest Path]
A Logical Algorithms implementation of Dijkstra's shortest path algorithm is
given in \cite{la} and in Example~\ref{ex:dijkstra}. A very similar
implementation in \chrrp{} is given in \cite{chrrp} and shown below.
\begin{Verbatim}[fontsize=\small,frame=single,commandchars=*\{\}]
    1 :: d1 @ source(V) ==> dist(V,0).
    1 :: d2 @ dist(V,D*vrbsubt{1}) \ dist(V,D*vrbsubt{2}) <=> D*vrbsubt{1} < D*vrbsubt{2} | true.
D + 2 :: d3 @ dist(V,D), e(V,C,U) ==> dist(U,D+C). 
\end{Verbatim}
Given a goal consisting of one \texttt{source}/1 constraint and $e$ 
\texttt{e}/3 constraints, the runtime complexity of this implementation is
$\mathcal{O}(e\log e)$. The analysis is essentially the same as the one for the
Logical Algorithms implementation as given in \cite{la}; see also 
Example~\ref{ex:dijkstra}.
\end{example}
\subsection{Comparison with the Logical Algorithms meta-complexity result}
\label{sec:new_meta:comparison_la}
In \cite{chrla}, we have presented a direct implementation of the Logical 
Algorithms language into CHR that satisfies the complexity requirements needed
for the Logical Algorithms meta-complexity result to hold. In this subsection,
we show that this implementation has become somewhat obsolete because we can
achieve the same result by combining the translation from Logical Algorithms to
\chrrp{} of Section~\ref{la-to-chr}, with the \chrrp{} implementation presented
in Section~\ref{sec:new_implementation}. We assume here that the comparison 
antecedents in Logical Algorithms programs are scheduled after the 
corresponding user-defined antecedents in the translation, and that the guards
on the mode indicators (these have the form $N\neq\texttt{p}$) are scheduled
right after the head to which they apply.
\begin{theorem}
The time complexity of Logical Algorithms programs executed by first 
translating them into \chrrp{} programs using the translation schema of 
Section~\ref{la-to-chr}, and then executing the resulting \chrrp{} program
using the implementation of Section~\ref{sec:new_implementation}, is
$\mathcal{O}(|\sigma_0|+P_s+(P_d+A_d)\cdot\log N)$ with $S_0$, $P_s$,
$P_d$, $A_d$ and $N$ as defined in Section~\ref{sec:la_and_chr:meta:la}.
\end{theorem}
\begin{proof}
The translation of a Logical Algorithms program $P$ consists of two parts as
defined in Section~\ref{la-to-chr}. The first part, denoted by $T_{S/D}(P)$,
contains for each user-defined predicate $a/n$ the following rules:
\begin{align*}
1 :: a_r(\bar{X},M)\ \backslash\ a(\bar{X}) &\iff M \neq \mathtt{n}\mid true\\
1 :: a_r(\bar{X},\mathtt{n}), a(\bar{X}) &\iff a_r(\bar{X},\mathtt{b})\\
2 :: a(\bar{X})&\iff a_r(\bar{X},\mathtt{p})\\
\vphantom{\stackrel{\displaystyle{m}}{T}}%
1 :: a_r(\bar{X},M)\ \backslash\ del(a(\bar{X}))&\iff M\neq\mathtt{p}\mid true\\
1 :: a_r(\bar{X},\mathtt{p}), del(a(\bar{X}))&\iff a_r(\bar{X},\mathtt{b})\\
2 :: del(a(\bar{X}))&\iff a_r(\bar{X},\mathtt{n})
\end{align*}
It is easy to see that for an initial goal containing no constraints of the
form $a_r(\bar{X},M)$ and since these are the only rules that assert such a 
constraint, in any state it holds that if $a_r(\bar{X},M_1)\#i_1$ and
$a_r(\bar{X},M_2)\#i_2$ are in the CHR constraint store, then $i_1=i_2$ and
$M_1=M_2$. This implies that the number of strong prefix firings for these 
rules is bounded by the number of assertions of $a(\bar{X})$ or 
$del(a(\bar{X}))$.

The second part of the translation, denoted by $T_R(P)$, contains for each
Logical Algorithms rule
\[r\ @\ p: A_1,\ldots,A_n\Rightarrow C\]
a set of rules
\[p+2::r_\rho\ @\ H\implies g_1,g_2\mid C'\]
as shown in the translation schema of 
Section~\ref{sec:la_and_chr:schema:rules}. Amongst these rules is one,
say $r_{\rho'}$, with a maximal number of heads, namely as many as
there are user-defined antecedents in $A_1,\ldots,A_n$. Because the (implicit
and explicit) guards on the mode indicators of the head constraints are 
scheduled as soon as they are decidable, and because the comparisons are
scheduled at corresponding places, it is easy to see that the number of strong
prefix firings of rule $r_{\rho'}$ is the same as the number of strong prefix
firings of Logical Algorithms rule $r$. The other $r_\rho$ rules are restricted
versions of $r_{\rho'}$ and therefore have at most as many strong prefix 
firings as $r_{\rho'}$.

The assertions with occurrences in dynamic priority rules are of the form
$a_r(\bar{X},\avar)$. 
The set and deletion semantics rules ensure that the number
of these assertions is the same in the original program and in its translation.
Finally, we note that the number of assertions with occurrences in a static
priority rule, $A_s$, is bounded by the number of assertions in the initial
goal $|\sigma_0|$ plus the number of rule firings times the maximal number of
body literals in any rule. Therefore, $A_s=\mathcal{O}(|\sigma_0|+P_s+P_d)$.
Now using our new meta-complexity result for \chrrp{} (Theorem~\ref{the:meta}),
we derive that the total runtime complexity of the translated program is 
$\mathcal{O}(|\sigma_0|+P_s+(P_d+A_d)\cdot\log N)$.
\end{proof}
\subsection{Comparison with the ``As Time Goes By'' approach}
\label{sec:new_meta:comparison_atgb}
In Section~\ref{sec:la_and_chr:meta:comparison} we already briefly compared
the Logical Algorithms meta-complexity theorem with the theorem given by
Fr\"uhwirth in \cite{atgb2}. In this subsection, we make the comparison
complete by also considering built-in constraints, using the new 
meta-complexity theorem presented in Section~\ref{sec:new_meta}. 

Let there be given a \chrrp{} program $P$ in which each rule has the same
(static) priority. Theorem 3 in \cite{chrrp} states that such a \chrrp{} 
program and its corresponding CHR program (which is found by removing the
rule priorities) have the same derivations. Therefore, such programs are
suitable for comparing the result of \cite{atgb2} with the result of 
Theorem~\ref{the:meta} in Section~\ref{sec:new_meta}. In 
Section~\ref{sec:la_and_chr:meta:comparison} we have already shown that
the number of strong prefix firings is 
$\mathcal{O}\left(D\cdot\sum_{r\in P}c^{n_r}_\mathit{max}\right)$ where
$D$ is the derivation length (i.e., the number of rule firings), and 
$c_\mathit{max}$
is the maximal number of CHR constraints in the store in any state. The number
of constraint assertions is $\mathcal{O}(c_\mathit{max}+D)$. 
If we assume that the 
initial goal does not contain any built-in constraints (as is done in 
\cite{atgb2}), then the number of built-in constraints is $\mathcal{O}(D)$. The
number of suspended prefix firings is bounded by 
$\mathcal{O}\left(\sum_{r\in P}c^{n_r}_\mathit{max}\right)$ in
any state and the number of suspended assertions by 
$\mathcal{O}(c_\mathit{max})$.
Now, filling in these parameters in the \chrrp{} meta-complexity result gives 
us that the total runtime complexity is
\begin{equation}\label{eq:comparison:ours}
\mathcal{O}\left((1+O_C)\cdot D\sum_{r\in P}(c^{n_r}_{max}\cdot O_{G_r})\right)
\end{equation}
where $O_C=\sum_{r\in P}(O_{C_r})$.
This formula strongly resembles the result of \cite{atgb2} which, assuming the
cost of head matching $O_{H_r}$ and adding and removing CHR constraints 
$O_{B_r}$ is constant, equals 
\begin{equation}\label{eq:comparison:atgb}
\mathcal{O}\left(D\sum_{r\in P}(c^{n_r}_{max}\cdot O_{G_r}+O_{C_r})\right)
\end{equation}
The difference lies in how built-in tell constraints are dealt with. In our
\chrrp{} implementation, as well as in any CHR implementation based on the
refined operational semantics of CHR, a built-in tell constraint causes the
constraints or matches whose variables are affected, to be reconsidered.%
\footnote{Which constraints are reactivated depends on the wake-up policy used
for the \textbf{Solve} transition, see also \cite[Section 5.4.2]{tom:phd}.} 
Because each individual (atomic) built-in constraint is dealt with separately,
this may cost more in total than the naive approach taken in \cite{atgb2} in
which after each rule firing, \emph{all} constraints or matches are 
reconsidered \emph{once}. 
So, while in certain rather exceptional cases, a naive approach to dealing with 
built-in tell constraints might in fact be better than the usual approach of
selective reactivation (as can be seen by comparing Formulas 
\eqref{eq:comparison:ours} and \eqref{eq:comparison:atgb}), in general we 
expect the latter approach to be an improvement over the naive one. Moreover,
in these exceptional cases, the meta-complexity theorem of \cite{atgb2} does
not apply to optimized CHR implementations like the K.U.Leuven CHR system, 
i.e., in these cases it does not overestimate the actual worst case time
complexity.

Noteworthy is that the approach of \cite{atgb2} only considers simplification
(and implicitly also simpagation) rules. This restriction is related to the
termination analysis which is used to find an upper bound on the number
of rule applications. However, if we can find such an upper bound by other
means, also propagation rules can be supported. For instance, the
termination analysis presented in \cite{termination4} can be used for this 
purpose.
\section{Conclusions}\label{sec:conclusions}
In this paper, we have investigated the relationship between the Logical
Algorithms language and Constraint Handling Rules. We have presented an 
elegant translation schema from Logical Algorithms to \chrrp{}: CHR extended
with user-definable rule priorities. The original program and its
translation are shown to be essentially weakly bisimilar. However, our current
\chrrp{} system \cite{compiling_chrrp} does not give the complexity guarantees
needed for the Logical Algorithms meta-complexity theorem to hold via this
translation. 

As a first step towards applying the Logical Algorithms meta-complexity result
to \chrrp{} programs, we have shown how a subclass of \chrrp{} can be
translated into Logical Algorithms. By using this translation, we can directly
apply the meta-complexity theorem for Logical Algorithms to the translated
\chrrp{} programs. A drawback is that the \chrrp{} programs that can be 
translated this way, are restricted to those that do not make use of an
underlying constraint solver. 

In order to remedy both the limitation that the translation from Logical 
Algorithms to \chrrp{} does not exhibit the required complexity when executing
translated Logical Algorithms programs 
using our \chrrp{} system, and the restriction of those
\chrrp{} programs that can be translated to Logical Algorithms and hence to
which the Logical Algorithms meta-complexity result can be applied, we have
proposed a new implementation for the complete \chrrp{} language that gives
strong complexity guarantees. The implementation is based on the high-level
implementation proposal of \cite{la} as well as on the scheduling data 
structure of \cite{mergeable_schedules}, and consists of the compilation of
\chrrp{} rules into (regular) CHR rules, combined with a scheduler that 
controls the execution. The implementation supports a new and accurate 
meta-complexity theorem for \chrrp{}. When combining the translation from
Logical Algorithms to \chrrp{} with the new implementation, the new 
meta-complexity theorem implies the Logical Algorithms meta-complexity result.
Moreover, it is shown that in general -- apart from some rather
exceptional cases, see Section~\ref{sec:new_meta:comparison_atgb} -- the new 
theorem is at least as accurate as the meta-complexity result for CHR given by
Fr\"uhwirth in \cite{atgb2}. This is illustrated on two non-trivial
examples, one of which contains both built-in constraints and propagation rules
and therefore cannot be analyzed using the Logical Algorithms approach or
Fr\"uhwirth's result.

\subsection{Related Work}
The time complexity of programs is in general expressed in terms of the 
number of elementary operations, e.g., the number of logical inferences in 
Prolog, function applications in a functional programming language, or rule
applications in a language such as CHR. However, while in most languages, these
elementary operations all take constant time, this is not the case in a 
language like CHR where each rule application results from a complex 
matching phase. 

In this work, we have made a mapping from the number of elementary operations 
(like prefix and rule firings or constraint assertions) to time complexity.
To the best of our knowledge, and apart from the results in \cite{la1,la2,la}
and \cite{atgb1,atgb2}, there is no other work with a similar goal. There are
many other formalisms though in which elementary operations take more than
constant time. One such formalism is term rewriting, as implemented by the 
Maude system \cite{maude} or the ACD term rewriting language \cite{acdtr}. It 
is known that AC matching, which is used by most of these languages, is 
NP-complete. Another formalism
is that of production rule systems like Drools \cite{drools} or Jess 
\cite{jess}. Production rules are in many ways similar 
to Constraint Handling Rules. However unlike CHR, these systems are not often
used as general purpose programming language, and therefore, algorithmic
complexity has never been much of a concern.
More work exists on the derivation of the number of elementary operations. In
the context of CHR, this mostly concerns the number of rule firings, which is
often derived as part of termination analysis 
\cite{termination1,termination2,termination3}.

Another related topic is that of space complexity, an issue that is not 
dealt with in this paper. In the context of CHR, the memory reuse techniques
developed in \cite{jon:memory_reuse} are crucial to achieve optimal space
complexity as is shown in \cite{jon:complexity:journal}. The latter also 
introduces a space complexity meta-theorem for CHR, stating that the space
complexity is $\mathcal{O}(D+p)$ where $D$ is the derivation length and $p$ is
the number of propagation rule firings (which takes into account the size of
the propagation history).

\subsection{Future Work}
For a previous version of this paper \cite{chrla}, we have made an actual
implementation for the Logical Algorithms language in CHR. This implementation
satisfies the complexity requirements needed for the Logical Algorithms 
meta-complexity theorem to hold, 
when executed using the K.U.Leuven CHR system on top of 
SWI-Prolog. However, the very large constant factors and the high memory
consumption makes that the system is not very 
useful in practice. Currently, we have
no running version of the alternative implementation for \chrrp{} presented in
Section~\ref{sec:new_implementation}. The reason is that this implementation
proposal is based on a similar approach as the Logical Algorithms one, and
in particular the more complicated scheduler is expected to be slow in 
practice. However, we do intend to 
investigate the advantages and disadvantages of a lazy RETE based matching
algorithm for CHR($^\mathrm{rp}$) compared to the LEAPS style matching that is
currently used by almost all systems. A simplified version of the scheduling
data structure of \cite{mergeable_schedules} which would offer less complexity
guarantees, but might be faster in the average case, could be used for this
purpose.

We have already mentioned in the related work discussion that a space
complexity result for our alternative implementation is currently lacking. The
RETE style matching we used is in general far from optimal as far as memory 
usage is concerned, in particular compared to LEAPS style matching as is used 
by most CHR systems. However, in the CHR context, built-in constraints may 
require maintaining a propagation history which in the worst case requires as
much memory as the alpha and beta memories in RETE matching. Therefore, it 
would be interesting to more formally compare both styles of matching in terms
of memory consumption in the context of CHR.

\section*{Acknowledgements}
The author would like to thank Tom Schrijvers, Bart Demoen and the anonymous
reviewers for their helpful and insightful comments. This research is funded
by a Ph.D. grant of the Institute for the Promotion of Innovation through
Science and Technology in Flanders (IWT-Vlaanderen).

\bibliographystyle{acmtrans}
\bibliography{paper}
\end{document}